\documentclass[aps,prc,nofootinbib, superscriptaddress]{revtex4-2}
\usepackage{graphicx}
\graphicspath{{figures/}}

\usepackage[dvipsnames]{xcolor}
\usepackage{amsmath, amsthm, amssymb, amsfonts}
\usepackage{slashed}
\usepackage{physics}
\usepackage{siunitx}
\usepackage{subcaption}
\usepackage{epsfig}
\usepackage{graphics}
\usepackage{slashed}
\usepackage{bbold}
\usepackage{empheq}
\usepackage{mathrsfs}
\usepackage{dcolumn}
\usepackage{slashed}

\usepackage{latexsym}
\usepackage{siunitx}
\usepackage{bm}
\usepackage{bbm}
\usepackage{microtype}
\usepackage{xspace}
\usepackage{placeins}

\usepackage{hyperref}
\hypersetup{
    colorlinks=true,
    allcolors = blue
    }

\usepackage[capitalise]{cleveref}

\usepackage{orcidlink}              
\usepackage{subcaption}
\usepackage{threeparttable}

\usepackage{multirow}

\usepackage{ulem}
\newcommand\redsout{\bgroup\markoverwith{\textcolor{red}{\rule[0.5ex]{2pt}{1.4pt}}}\ULon}

\renewcommand{\emph}[1]{\textit{#1}}

\newcommand{\Galilean}{\overset{\leftrightarrow}{\nabla}}

\newcommand{\calO}{\ensuremath{\mathcal{O}}}
\newcommand{\calD}{\ensuremath{\mathcal{D}}}

\newcommand{\calC}{\ensuremath{\mathcal{C}}}
\newcommand{\calA}{\ensuremath{\mathcal{A}}}
\newcommand{\calL}{\ensuremath{\mathcal{L}}}

\newcommand{\calZ}{\ensuremath{\mathcal{Z}}}

\newcommand{\hatr}{\ensuremath{\Hat{\vb{r}}}}

\newcommand{\museven}{\ensuremath{\mu}_L\xspace}

\newcommand{\mut}{\ensuremath{\mu}_T\xspace}
\newcommand{\muh}{\ensuremath{\mu}_H\xspace}
\newcommand{\mub}{\ensuremath{\mu}_B\xspace}

\newcommand{\fm}{\ensuremath{\mathrm{fm}}}

\newcommand{\Liseven}{\ensuremath{{}^7\text{Li}}\xspace}
\newcommand{\Be}{\ensuremath{{}^7\text{Be}}\xspace}
\newcommand{\Hethree}{\ensuremath{h}\xspace}

\newcommand{\Za}{\ensuremath{Z_\alpha}\xspace}
\newcommand{\Zhe}{\ensuremath{Z_h}\xspace}
\newcommand{\Zt}{\ensuremath{Z_t}\xspace}

\newcommand{\triton}{\ensuremath{t}\xspace}
\newcommand{\ZLisev}{\ensuremath{Z_{L}}\xspace}
\newcommand{\ZBe}{\ensuremath{Z_{B}}\xspace}

\newcommand{\Cpq}{\ensuremath{\calC_{3/2}}}
\newcommand{\Cpd}{\ensuremath{\calC_{1/2}}}

\newcommand{\LMone}{\ensuremath{L_{\rm M1}}}

\newcommand{\msev}{\ensuremath{m_{L}}\xspace}
\newcommand{\mbe}{\ensuremath{m_{B}}\xspace}
\newcommand{\mhthree}{\ensuremath{m_{H}}\xspace}
\newcommand{\mtthree}{\ensuremath{m_{T}}\xspace}

\newcommand{\malpha}{\ensuremath{m_{\alpha}}\xspace}

\newcommand{\nnlo}{\ensuremath{\rm N^2LO}\xspace}

\newcommand{\quadlisev}{\ensuremath{\mathcal{Q}_s^{(\Liseven)}}\xspace}

\newcommand{\quadbe}{\ensuremath{\mathcal{Q}_s^{(\Be)}}\xspace}

\newcommand{\Blisev}{\ensuremath{B_{\Liseven}}}
\newcommand{\Bbe}{\ensuremath{B_{\Be}}}

\newcommand{\calZBlisevq}{\ensuremath{\calZ^{(\Liseven)}_q}}
\newcommand{\calZBlisevd}{\ensuremath{\calZ^{(\Liseven)}_d}}

\newcommand{\Dpwave}{\ensuremath{D_\Pi}\xspace}

\newcommand{\Pquartet}{\ensuremath{{}^2P_{\frac{3}{2}}}\xspace}
\newcommand{\Pdoublet}{\ensuremath{{}^2P_{\frac{1}{2}}}\xspace}

\begin{document}
\title{Electromagnetic form factors of \Liseven and \Be in cluster effective field theory}
\author{Son T. Nguyen\orcidlink{0000-0002-6104-7035}}
\email{snguyen@wlu.edu}

\affiliation{Department of Physics and Engineering, Washington and Lee University, Lexington, Virginia 24450, USA
}
\affiliation{ Department of Physics, Box 90305, Duke University, Durham, North Carolina 27708, USA}

\begin{abstract}
 Effective field theory (EFT) provides a powerful model-independent theoretical framework for illuminating complicated interactions across a wide range of physics areas and subfields. In this work, we study the electromagnetic form factors of Lithium-7  and Beryllium-7 at low momentum transfers in cluster EFT, treating the nuclei as triton-Helium-4 and helion-Helium-4 bound states. We demonstrate how to directly extract the asymptotic normalization coefficients from the experimental measurements of charge radii, and use them  to calculate  several  electromagnetic observables of these nuclei. 
 
\end{abstract}

\maketitle

\section{Introduction}
The electromagnetic (EM) properties of lithium and beryllium isotopes have attracted much interest for many years. In particular, the production rates of these nuclei are an important input to the Big Bang nucleosynthesis model \cite{osti_1802279, Asplund:2005yt, RevModPhys.88.015004}. They also provide a crucial test bed for nuclear structure calculations for light nuclei consisting of more than four nucleons. Many different theoretical approaches using various degrees of freedom have been proposed over the years, such as nuclear potential models \cite{kajino1984electromagnetic, BOUTEN1968385,kajino19863he, Mason:2008ka,buck1985cluster,Mohr:2009uz}, effective range expansion \cite{Igamov:2007svc}, effective field theories (EFTs) \cite{Higa:2016igc, Zhang:2019odg, Poudel:2021mii}, and, most recently \textit{ab initio} calculations \cite{Dohet-Eraly:2015ooa, Vorabbi:2019imi, PhysRevC.100.024304,hebborn2022ab,Atkinson:2024zrm, Henderson:2019ubp,Shen:2024qzi}.  

The asymptotic normalization coefficient (ANC) is a fundamental quantity in nuclear physics, providing a direct measure of the amplitude of the bound state wavefunction in the asymptotic region. It provides a powerful way to determine reactions involving weakly bound systems, such as radiative capture and transfer reactions, where it enables determining reaction rates without reliance on uncertain optical potentials. In particular, the ANC plays a vital role in nuclear astrophysics, offering a way to extrapolate $S$-factors for direct capture reactions to the low-energy regime relevant to stellar environments \cite{mukhamedzhanov2022asymptotic}. In this work, we study the EM form factors of Lithium-7 (\Liseven) and Beryllium-7 (\Be)  nuclei using cluster EFT for low-momentum transfer up to 1~\unit{fm^{-1}}. Their low-$q$ behavior is dominated by the asymptotic region of the bound-state wave function. The same region also governs radiative capture at astrophysical energies, where the relative momenta, $\lesssim0.2$~\unit{fm^{-1}}, are well within this range. Recent four-body microscopic calculations, in which the $\alpha$ particle is treated as a cluster while the formation of the triton (helion) is not assumed but emerges from $\alpha+n+n+p$ and $\alpha+p+p+n$ descriptions, find
the $\alpha+t$ and $\alpha+\Hethree$ configurations to dominate the low-lying
states of $^7$Li and $^7$Be respectively, with spectroscopic factors close to
unity~\cite{Tao:2025ukl}. We focus on benchmarking the value of ANCs for the reactions, $\alpha + t \to$\Liseven and $\alpha + \Hethree \to$\Be, where $t, h$ stand for triton and helion, respectively. 

Cluster EFT gives an effective description of nuclei consisting of two or more bound clusters of nucleons. Since it involves minimal assumptions, we can obtain some universal relations that complement the {\it ab initio} approaches and experiments.  Within the applicable range, the clusters are considered point-like particles at leading order (LO), with their finite size entering at higher orders. A similarly motivated method is halo EFT, which treats halo nuclei as composites of one cluster and then one or more nucleons (see, e.g., Refs.~\cite{Hammer:2017tjm, Hammer:2019poc,ando2021cluster} for reviews). This framework has been successfully applied to study $\alpha-\alpha$ scattering, the physics of halo nuclei such as one- (e.g., $^{11}$Be, $^{15}$C, $^{19}$C) and two-neutron halo configurations (e.g. $^{11}$Li, $^{14}$Li, $^{22}$C), one-proton halo configurations (e.g., $^{8}$B), and $\alpha$-halo configurations (e.g., $^{16}$O). In fact, $\alpha$ is very stable, and the single-nucleon separation energies and the excited states are around 20 MeV.

Cluster EFT takes advantage of the separation of scales existing in the system to provide a systematic and model-independent expansion in a ratio $Q\equiv k_{\rm low}/\Lambda$, where $ k_{\rm low}$ is the typical low-momentum scale and $\Lambda$ is the momentum scale at which the theory breaks down. Cluster EFT provides a significant simplification by reducing these complex problems to an effective two-body problem. Furthermore, the Coulomb interaction between two clusters can be solved analytically. In this work, we will focus on the weakly bound states of two nuclear clusters, so it is convenient to utilize a dimer field formalism already outlined in Refs.~\cite{PhysRevC.89.014325,Ryberg:2015lea}. 

Another key benefit of EFTs lies in their ability to naturally estimate the size of errors due to neglected higher-energy effects. The breakdown scale $\Lambda$ of the cluster EFT typically corresponds to the momentum associated with the excitation of internal degrees of freedom within the clusters or the energy scale at which new relevant degrees of freedom, such as excited states or substructure effects, become relevant.  The typical momentum
scale $k_{\rm low}$ is the larger of the binding momentum and the momentum transfer. The theoretical error of an EFT calculation at the order $\calO(Q^n)$ in the power counting is of order $(k_{\rm low}/\Lambda)^{n+1}$. Whenever feasible, we propagate the uncertainties of input parameters to the corresponding low-energy observables.

Lithium-7 can be treated as a two-cluster nucleus composed of an alpha particle ($\alpha$) and a triton ($t$). The $\alpha$ nuclei excitation energy is $E_\alpha^*\approx 20$ MeV. The binding energies of its ground state ($J^\pi = 3/2^-$) and first excited state ($J^\pi = 1/2^-$) are 2.467 MeV and 1.989 MeV, respectively, relative to the $\alpha + t$ threshold.  These binding energies are below the triton breakup energy at  $S_n(t)=6.26$ MeV, which corresponds to a momentum of 142~MeV. As discussed in \cref{sec:lithium_seven}, the empirically validated breakdown scale is $\Lambda\approx200$~MeV. The relatively small binding energies in comparison to this scale make cluster EFT a suitable framework for providing controlled and model-independent calculations of \Liseven's properties, such as its EM form factors. The treatment of EM interactions in a dimer formalism follows the
approach developed for the two-nucleon system in pionless
EFT~\cite{Chen:1999tn,Chen:1999vd, RUPAK2000405,PhysRevC.60.065205}.
The \Liseven nucleus has been the subject of extensive theoretical and experimental studies, including approaches using $\delta$-shell potentials~\cite{PhysRevC.100.054307}, highlighting its importance as a benchmark for nuclear structure and reaction models.

Similarly, Beryllium-7 can be characterized as a two-cluster nucleus composed of an alpha particle and a helion (\Hethree). Its ground state ($J^\pi = 3/2^-$) and first excited state ($J^\pi = 1/2^-$) are bound at 1.586 MeV and 1.157 MeV, respectively, below $\alpha + \Hethree$ threshold \cite{TILLEY20023}. The helion's breakup energy is 5.49~MeV, placing the binding energies of \Be well within the regime where cluster EFT is applicable. This model-independent approach provides a robust framework for studying the EM properties of \Be, such as its magnetic form factor and quadrupole moment, which have not yet been measured.

 In this work, we estimate the ANCs using measured charge radii data and then use them as inputs to make predictions of electromagnetic form factors. An important question is how precise the ANCs must be in order for such predictions to be astrophysically meaningful. At low energies many radiative capture reactions of astrophysical interest are peripheral so that the cross section and the corresponding zero-energy S-factor, $S(0)$ is proportional to the square of the ANC. The uncertainty in $S(0)$ is typically of order 5–10\% (see, e.g., Ref.~\cite{RevModPhys.83.195}), which requires the ANC to be determined to a few percent.  A notable example is the study of the radiative ${}^{3}{\rm He}(\alpha,\gamma)\Be$ reaction within Halo EFT \cite{Higa:2016igc}, which demonstrates how $S(0)$ can be extracted from scattering data. As pointed out by those authors, most available elastic scattering data lie at the high-energy end of the EFT’s validity range, leading to instabilities in fitting the low-energy parameters. The difficulty is compounded by
the fact that the ANC depends on the full set of effective-range parameters,
not on the scattering length alone~\cite{Ryberg:2015lea}. Electromagnetic
observables offer a complementary route because they are measured with high precision
and probe the system near zero momentum transfer. The ANC can be fitted
directly to data without extracting each effective-range parameter
individually~\cite{Ryberg:2015lea}. Assessed against this requirement, the extraction presented below yields an ANC uncertainty of $12\%$ if the breakdown scale is identified with the triton-breakup momentum, corresponding to $25\%$ in $S(0)$. However, we show that if the uncertainty is
estimated from data instead either
from the empirically determined breakdown scale of Ref.~\cite{Poudel:2021mii},
or from the benchmarks against the measured $^7$Li quadrupole moment and
transition probability,  the uncertainties of
ANCs reduce to the few-percent level ($5\%$ for \Liseven and $3\%$ for \Be) and bring $S(0)$ within the target.

Our work is organized as follows. \Cref{sec:lithium_seven} focuses on the \Liseven nucleus. We introduce our effective Lagrangian and describe how we derive relationships between the theory's low-energy coefficients (LECs) and the ANC. We then fix the ANC using the measured charge radius, and use the charge form factor, the quadrupole moment, and the $B({\rm E2})$ transition strength as benchmarks to verify the breakdown scale and our uncertainty estimates. The magnetic and longitudinal form factors are then predictions of the theory. In \cref{sec:beseven} we apply the same procedure to the \Be nucleus. Finally, we present our conclusions in \cref{sec:conlusion}.

\section{Lithium-7 \label{sec:lithium_seven}}
The binding energy of the ground state of \Liseven $(J^{\pi}=3/2^{-})$ is $B_{\Liseven}=2.47$ MeV, corresponding to a binding momentum of $\gamma_{q}=89$ MeV or $0.451$ \unit{\fm^{-1}} below the threshold of $\alpha +t$. The binding energy of the first excited state $(J^{\pi}=1/2^{-})$ is $B_{\Liseven^*}=1.99$~MeV, corresponding to a binding momentum of $\gamma_{d}=80.0$ MeV or $0.406$ \unit{\fm^{-1}}. As discussed earlier, $\alpha$ is very stable, whereas the neutron separation energy of the triton is about $S_n(t)=6.26$ MeV, which would naively set the breakdown scale of the theory at $\Lambda\approx 142$ MeV. We note that the breakdown scale can also be determined directly from data. Ref.~\cite{Poudel:2021mii} reconstructs the order-by-order expansion coefficients for the mirror system $\alpha+h$ and finds that the hard scale is instead associated with the first excitation of the $\alpha$
particle, giving $\Lambda \approx 200$~MeV. No such analysis exists for $\alpha+t$, and repeating it is beyond the scope
of this paper. We therefore apply a more limited test, asking only whether
the coefficient of the first omitted order beyond N$^2$LO is of natural size.
Within that test, applied to the $^7$Li charge form factor and to the
parameter-free predictions for $\mu$, $\mathcal{Q}_s$ and $B({\rm E2})$, the
coefficients are of order unity for $\Lambda \approx 200$~MeV and
unnaturally small for $\Lambda \approx 142$~MeV. Our results are therefore
consistent with the larger scale in the $\alpha+t$ channel as well, although
they do not by themselves establish it.

The effective Lagrangian was motivated by the one developed in Refs.~\cite{Higa:2016igc, Zhang:2019odg} for the $\alpha + \Hethree\to \Be+\gamma$ reaction, which reads
\begin{equation}
\begin{aligned}
  \calL =&~ t^\dagger_a\left(i\calD_0 + \frac{\boldsymbol{\calD}^2}{2m_t}\right)t_a + \phi^\dagger\left(i\calD_0 + \frac{\boldsymbol{\calD}^2}{2m_\alpha}\right)\phi\\
  & + \Pi^\dagger_{x}\left[\Delta_\Pi +\nu_q\left(i\calD_0  +\frac{\boldsymbol{\calD}^2}{2\msev}\right)\right]\Pi_{x}  + \Xi^\dagger_a\left[\Delta_\Xi +\nu_d\left(i\calD_0  +\frac{\boldsymbol{\calD}^2}{2\msev}\right)\right]\Xi_a  \\
  & + g_q\Pi^\dagger_{x} \left(S^\dagger_{i}\right)_{xa}\left( t_a i\Galilean_i \phi +\text{h.c.}\right) + g_d\Xi^\dagger_a \left[\sigma_i\right]_{ab}\left( t_bi\Galilean_i \phi +\text{h.c.}\right), 
\end{aligned}  
\label{eq:lithium_sev_scattering}
\end{equation}    
where $\Pi$ and $\Xi$ ($x=1,2,3,4$, $i,j=1,2,3$ and $a,b=1,2$) are auxiliary fields associated with the \Pquartet and \Pdoublet channels, respectively. $\Delta_\Pi,\Delta_\Xi, g_q, g_d, \nu_q$, and $\nu_d$ are undetermined LECs, in which $\Delta_\Pi$ and $\Delta_\Xi$ are interpreted as the residual mass of the $\Pi$ and $\Xi$ auxiliary fields. The $S_i$’s are the $2 \times 4$ spin-transition matrices (see Appendix~\ref{app:em_current}) connecting states with total angular momentum $j = 1/2$ and $j = 3/2$. In addition, the total mass of the dimer field is $\msev=m_t+m_\alpha$, where $m_t$ is the triton mass.  Here, the Galilean derivative $\Galilean$ is defined as
\begin{equation}
 t ~\Galilean_i ~\phi = \frac{m_t}{\msev} t\Vec{\nabla}_i\phi  - \frac{m_\alpha}{\msev}\left(\Vec{\nabla}_it\right) \phi.
\end{equation}
Electromagnetic charge interactions are included via minimal substitution in which the gradient is replaced by the covariant derivative $\calD_\mu=\nabla_\mu+i\hat{Q} eA_\mu$, where $\hat{Q}$ is the charge operator.
The dressed propagator for the $\Pi$-dimer  is 
\begin{equation}
    i\Dpwave(E) =
    \frac{i\delta_{xy}}{\Delta_\Pi +\nu_q (E+i\epsilon) +\Sigma_{q}(E)},
\end{equation}
where $\Sigma_{q}(E)$ is the irreducible self-energy, which was already outlined in Ref.~\cite{Kong:1999sf}. A detailed derivation of this quantity, including its regularization in the power divergence subtraction scheme \cite{Kaplan_KSW_1996,Kaplan_KSW_1998a,Kaplan_KSW_1998b}, was also presented in Ref.~\cite{Stellin:2020gst}. The divergences of $\Sigma_{q}(E)$  and its energy derivative are absorbed into the combinations $\Delta_\Pi/g_q^2$ and $\nu_\Pi/g_q^2$. The spin-quartet $P$-wave scattering amplitude in the c.o.m frame is 
\begin{equation}
    i\calA_q =  -6ig_q^2\Dpwave(E)C_{\eta,1}^2e^{2i\sigma_1}\vb{p}\cdot\vb{p}'.
\end{equation}
In the above equation, $\sigma_1$ is the $\ell =1$ Coulomb phase shift and $C_{\eta, 1}$ is Sommerfeld factor representing the probability
of the two charge nuclei at the zero-separation, given by
\begin{equation}
    C_{\eta, 1} = \frac{1}{3}e^{-\pi\eta/2}|\Gamma(2+i\eta)|. 
\end{equation}
We can calculate the residue of the renormalized $\Pi$-dimer propagator  via
\begin{equation}
\frac{1}{\calZBlisevq}   = \frac{\dd }{\dd E}\left[\frac{1}{\Dpwave(E)}\right]\Bigg|_{E=-\Blisev} = \nu_q +\Sigma'_{q}(-\Blisev).
\label{eq:LSZ_lisev}
\end{equation}
For $P$-waves, an exact relationship between the ANC and the wave function renormalization coefficient is given by \cite{Higa:2016igc}
\begin{equation}
    \calZBlisevq = \frac{\pi }{2g_q^2m_r^2\gamma_{q}^2}\frac{1}{\Gamma(2+\eta_{q})^2} \left[\Cpq\right]^2,
\end{equation}
where $\eta_{q}=\alpha_{\rm em} \Za\Zt m_r/\gamma_{q}$ and $m_r$ is the reduced mass of the $\alpha$-$t$ system. Similarly, we can obtain a relationship between the residue of the renormalized $\Xi$-dimer propagator and the ANC as
\begin{equation}
     \calZBlisevd = \frac{\pi }{2g_d^2m_r^2\gamma_{d}^2}\frac{1}{\Gamma(2+\eta_{d})^2} \left[\Cpd\right]^2.
\end{equation}
Here $\eta_{d}=\alpha_{\rm em} \Za\Zt m_r/\gamma_{d}$. The non-relativistic expansion of the matrix element of the EM current is given as 
\begin{align}
    \langle \vb{p}', x|J^0_{\rm em}|\vb{p},y\rangle &= e\left[F^{(\Liseven)}_{\rm C0}(q)\delta_{xy} + \frac{1}{2(\msev)^2} F^{(\Liseven)}_{\rm C2}(q)\left[S_i^\dagger S_j\right]_{xy}\left(q_iq_j-\frac{1}{3}\delta_{ij}q^2\right)  \right],\\
    \langle \vb{p}', x|J^k_{\rm em}|\vb{p},y\rangle &= \frac{e}{2\msev}F^{(\Liseven)}_{\rm M1}(q)\left[S_i^\dagger\right]_{xa}\left[i\left(\boldsymbol{\sigma}\times \vb{q}\right)^k\right]_{ab}\left[S_i\right]_{by}\ ,
    \end{align}
where $\vb{q}=\vb{p}'-\vb{p}$ is the momentum transfer (see Appendix.~\ref{app:em_current} for more details). 
We also want to obtain the non-relativistic expression of the EM transition matrix element from \Liseven ($3/2^-$) to $\Liseven^*$ ($1/2^-$). We ignore the mass difference between  the two states, which is only 0.478 MeV.  Adopting the Jones and Scadron's convention \cite{Jones:1972ky}, the matrix element of $J^0_{\rm em}(0)$ in the Breit frame reduces to
\begin{equation}
    \langle \vb{p}', a|J^0_{\rm em}|\vb{p},x\rangle = \frac{e}{2(\msev)^2}F'_{\rm C2}(q)(\va*{\sigma}\cdot\va{q})_{ab}(\va{S}\cdot\va{q})_{bx}.
\end{equation}
Electromagnetic moments are related to these form factors at the real photon point $q=0$.
\begin{align}
F^{(\Liseven)}_{\rm C0} (0) &= \ZLisev,\\
 \frac{1}{2\msev} F^{(\Liseven)}_{\rm M1} (0) &= \mu^{(\Liseven)},\\
    \frac{1}{(\msev)^2}F^{(\Liseven)}_{\rm C2}(0) &=
\quadlisev,\\
\left[\frac{F'^{(\Liseven)}_{\rm C2}(0)}{(\msev)^2}\right]^2 &= \frac{16\pi}{25}B({\rm E2}), 
\end{align}
where $\ZLisev=3$, $\museven=3.256427\mu_N$ \cite{TILLEY20023}, $\quadlisev = -4.00(3)$ \unit{\fm^2} \cite{STONE20161}, and $B({\rm E2})$ is known as the reduced transition probability.
In the next section, we will use the Lagrangian in \cref{eq:lithium_sev_scattering} to calculate EM form factors and then determine $\Cpq$ and $\Cpd$ from available measurements of \Liseven's EM moments.

\subsection{\Liseven's charge form factor}

\begin{figure}[ht]
    \centering
    \includegraphics[width= .7\linewidth]{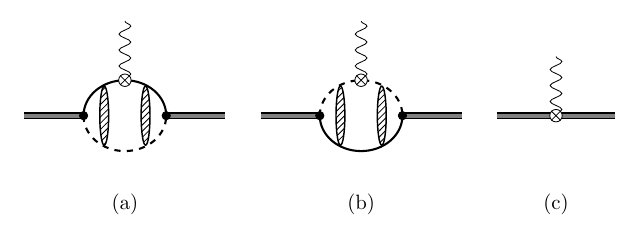}
    \caption{Diagrams for charge form factor of  \Liseven. The solid line stands for the $\alpha$ field, the dashed line can represent either the
triton or the helion, and the double line for the dressed propagator.  The wavy line illustrates a $A_0$ photon coupling to either constituents of the nuclear cluster or the dimer. They appear at LO in the power counting. These same diagrams
arise for \Be cluster. }
    \label{fig:E_form_factor}
\end{figure}

The LO diagrams that contribute to the \Liseven charge form factor in the
cluster EFT are shown in \cref{fig:E_form_factor}. In addition to diagrams where the photon couples to each constituent nuclei, there
are also couplings to the auxiliary field obtained by gauging the Lagrange density in \cref{eq:lithium_sev_scattering}. The irreducible three-point function shown in \cref{fig:E_form_factor}a yields
\begin{equation}
\begin{aligned}
    -i\Gamma^{(a)}_{\rm C} &=ig_q^2\Za \int\frac{\dd^4k_1\dd^4k_2\dd^4k_3}{(2\pi)^{12}} \left(\vb{k}_3-\frac{(1-f)\vb{q}}{2}\right)_i\left[S^\dagger_{i}S_j\right]_{xy}iS_\alpha\left(E_3, \vb{k}_3\right)iS_t\left(E-E_3,\frac{\vb{q}}{2}-\vb{k}_3\right)\\
    &\quad\times i\chi\left(\vb{k}_3-\frac{f\vb{q}}{2};\vb{k}_2-\frac{f\vb{q}}{2},-\Blisev\right)iS_t(E_2,\vb{k}_2) \left(\vb{k}_1+\frac{(1-f)\vb{q}}{2}\right)_j\\
    &\quad \times iS_\alpha\left(E-E_2,-\vb{k}_2+\frac{\vb{q}}{2}\right)iS_\alpha\left(E-E_2,-\vb{k}_2-\frac{\vb{q}}{2}\right)\\
    &\quad\times i\chi\left(\vb{k}_2+\frac{f\vb{q}}{2};\vb{k}_1+\frac{f\vb{q}}{2} ;-\Blisev\right) iS_\alpha\left(E_1, \vb{k}_1\right)iS_t\left(E-E_1,-\frac{\vb{q}}{2}-\vb{k}_1\right),    \end{aligned}
    \label{eq:charge_FF_Lseven}
\end{equation}    
where $f=m_t/\msev$, $S_\alpha$ and $S_t$ are the alpha and triton field propagators, and $\chi$ is the Coulomb four-point
function (see, e.g., Ref.~\cite{Ryberg:2015lea}). The total energy is given by $E=-\Blisev + q^2/8\msev$. We first evaluate the energy
integrals in \cref{eq:charge_FF_Lseven} using the residue theorem. This leads to
\begin{equation}
\begin{aligned}
    -i\Gamma^{(a)}_{\rm C} &=   ig_q^2 \Za \int \frac{\dd^3 \vb{k}_1\dd^3 \vb{k}_2\dd^3 \vb{k}_3}{(2\pi)^9} \langle \vb{k}_3|G_C(-\Blisev)|\vb{k}_2-f\vb{q}/2\rangle\langle \vb{k}_2+f\vb{q}/2|G_C(-\Blisev)|\vb{k}_1\rangle \left(\vb{k}_3\right)_i\left[S^\dagger_{i}S_j\right]_{xy} \left(\vb{k}_1\right)_j.
\end{aligned}
\label{eq:charge_FF_Lseven_simplified}
\end{equation}
We adopt the notation in \cite{Ryberg:2015lea}, which defined, $S_{\rm tot}(E,\vb{k})=i(E-\vb{k}^2/2m_r+i\epsilon)^{-1}$. The Coulomb Green’s function $G_C$ in the above equation is given by
\begin{equation}
    \langle \vb{k} | G_C(E)|\vb{p}\rangle=-S_{\rm tot}(E,\vb{k})\chi(\vb{k},\vb{p};E) S_{\rm tot}(E,\vb{p}).
\end{equation}
The integral in \cref{eq:charge_FF_Lseven_simplified} is better evaluated in coordinate space. Performing a Fourier transform on each of the
momentum-space bras and kets yields the following,
\begin{equation}
\begin{aligned}
    -i\Gamma^{(a)}_{\rm C} 
    &=   ig_q^2 \Za \int \frac{\dd^3 \vb{k}_1\dd^3 \vb{k}_3}{(2\pi)^6}\ \dd^3\vb{r}\ e^{-if\vb{q}\cdot\vb{r}} \langle \vb{k}_3|G_C(-\Blisev)|\vb{r})(\vb{r}|G_C(-\Blisev)|\vb{k}_1\rangle \left(\vb{k}_3\right)_i\left[S^\dagger_{i}S_j\right]_{xy} \left(\vb{k}_1\right)_j
\end{aligned}
\label{eq:charge_FF_Lseven_simplified_II}
\end{equation}
Using the identity (see Appendix C of Ref.~\cite{Ryberg:2014exa} for detailed derivation)
\begin{equation}
    \int\frac{\dd^3\vb{k}}{(2\pi)^3}\vb{k}\langle\vb{k}|G_C(E)|\vb{r})=3i\ \hatr\lim_{\vb{r}'\to 0}\left[\frac{G_C^{(1)}(-B,\vb{r},\vb{r}')}{r'}\right],
\end{equation}
where the limit is discussed in \cref{app:Green_partial_wave}.
Thus, we obtain
\begin{equation}
    -i\Gamma^{(a)}_{\rm C} = ig_q^2\Za\frac{m_r^2\gamma_{q}^2}{4\pi^2}\Gamma\left(2+\eta_{q}\right)^2 \int_0^\infty \dd r\left[S^\dagger_{i}S_j\right]_{xy}\left[\frac{4\pi}{3}\delta_{ij} j_0(fqr) -4\pi j_2(fqr)\left(\hat{q}_{i}\hat{q}_{j}-\frac{1}{3}\delta_{ij}\right)\right]\left|W_{-\eta_{q},\frac{3}{2}}(2\gamma_{q}r)\right|^2,
    \label{eq:three_point_E_form_factor_Lisev}
\end{equation}    
where $j_0(x)$ and $j_2(x)$ are the spherical Bessel functions and $W_{\kappa,\mu}(z)$ is the Whittaker W function \cite[\href{https://dlmf.nist.gov/13.14.E3}{(13.14.E3)}]{NIST:DLMF}. The first term contributes to the charge form factor, while the second term contributes to the quadrupole form factor. The tree-level diagram is given by
\begin{equation}
    -i\Gamma^{(c)}_{\rm C} = -i\nu_q \ZLisev.
    \label{eq:Lisev_charge_LO_constant}
\end{equation}
First we want to check the normalization of the charge form factor
\begin{equation}
    F^{(\Liseven)}_{\rm C0}(q)=\calZBlisevq\left( \Gamma^{(a)}_{\rm C}+\Gamma^{(b)}_{\rm C}+\Gamma^{(c)}_{C}\right). 
\end{equation}
We can show that the form factor is properly normalized to $F^{(\Liseven)}_{\rm C0}(q)\to \ZLisev$ as $q\to0$. Indeed, at $q=0$, 
\begin{align}
    \Gamma^{(a)}_{\rm C}(0) & =  g_q^2\Za \int\frac{\dd^3\vb{k}_3 \dd^3\vb{k}_1}{(2\pi)^{6}}\dd^3\vb{r} \langle\vb{k}_3|G_C(-\Blisev)|\vb{r}) \,(\vb{r}|G_C(-\Blisev)|\vb{k}_1\rangle\left(\vb{k}_{3}\right)_i\left[S^\dagger_{i}S_j\right]_{xy}\left(\vb{k}_{1}\right)_j \nonumber \\
    & = \Za \Sigma'_{q}(-\Blisev).
    \label{eq:charge_LO_diagram_alpha}
\end{align}
On the other hand, we can transform the above integral to position space and obtain
\begin{align}
    \Gamma^{(a)}_{\rm C}(0) 
    &=\Za\frac{g_q^2m^2_r\gamma_{q}^2}{3\pi}\Gamma\left(2+\eta_{q}\right)^2 \int_0^\infty \dd r ~\left|W_{-\eta_{q},\frac{3}{2}}(2\gamma_{q}r)\right|^2.
\end{align}
Thus, we can show that
\begin{equation}
  \Sigma'_{q}(-\Blisev) = \frac{g_q^2m^2_r\gamma_{q}^2}{3\pi}\Gamma\left(2+\eta_{q}\right)^2\int_0^\infty \dd r ~\left|W_{-\eta_{q},\frac{3}{2}}(2\gamma_{q}r)\right|^2.
  \label{eq:divergence_of_threepoint}
\end{equation}
Similarly,
\begin{align}
    \Gamma^{(b)}_{\rm C}(0) 
    & = \Zt \Sigma'_{q}(-\Blisev).
        \label{eq:charge_LO_diagram_t}
    \end{align}
Combining \cref{eq:LSZ_lisev,eq:Lisev_charge_LO_constant,eq:charge_LO_diagram_alpha,eq:charge_LO_diagram_t} we obtain $F^{(\Liseven)}_{\rm C0}=\ZLisev$. At LO, there is no free parameter.
At next-to-next-to leading order (N$^2$LO), there are contributions from the finite-size effects of $\alpha$ and $t$ and a new term. To specify, the form factor of $\alpha$ is given by a gauge invariant operator of the form, $-\Za \phi^\dagger\nabla^2A_0\phi$.\footnote{A more general form is given by $-\Za \phi^\dagger\left(\nabla^2A_0-\partial_0\boldsymbol{\nabla}\cdot \vb{A}\right)\phi$. We have used the Coulomb gauge ($\boldsymbol{\nabla}\cdot \vb{A}=0$) to simplify the expression.} For a small momentum transfer window, the resulting charge form factor  of the $\alpha$ is
\begin{equation}
   F_{\rm C}^{(\alpha)}(q) = 1-\frac{\langle r^2_C\rangle_\alpha}{6}q^2+\calO(q^4), 
\end{equation}
where  $\sqrt{\langle r^2_C\rangle_\alpha} =1.67824(83)$ fm is the charge radius of $\alpha$ \cite{Krauth:2021foz}. The triton charge form factors and charge radii can be similarly evaluated  as
\begin{equation}
   F_{\rm C}^{(t)}(q) = 1-\frac{\langle r^2_C\rangle_t}{6}q^2+\calO(q^4). 
\end{equation}
Here $\sqrt{\langle r^2_C\rangle_d} =2.127 78(27)$ fm is the charge radius of the deuteron \cite{RevModPhys.93.025010}.
In addition, there is a local short-range operator that comes in with an undetermined short-range parameter, $w_q$,
\begin{equation}
  \mathcal{L} = -\ZLisev\,\frac{w_q}{6}\,
                 \Pi^\dagger \nabla^2 A_0\, \Pi .
\end{equation}
Thus, the charge form factor of \Liseven up to and including N$^2$LO is given by
\begin{align}
    F^{(\Liseven)}_{\rm C0}(q) =&~ \frac{\Za}{6} [\Cpq]^2 \int_0^\infty \dd r \left[j_0(fqr)-1\right]\left|W_{-\eta_{q},\frac{3}{2}}(2\gamma_{q}r)\right|^2\left(1-\frac{\langle r_C^2\rangle_\alpha}{6}q^2\right)\nonumber\\
    &+\frac{\Zt}{6} [\Cpq]^2\int_0^\infty \dd r \left\{j_0\left[(1-f)qr\right]-1\right\}\left|W_{-\eta_{q},\frac{3}{2}}(2\gamma_{q}r)\right|^2\left(1-\frac{\langle r_C^2\rangle_{\triton}}{6}q^2\right)\nonumber\\
&+\calZBlisevq\ZLisev\left[\nu_q+\Sigma'_{q}(-\Blisev)\right] -\frac{q^2}{6}\calZBlisevq\left[\Xi_q + \left(\Za \langle r_C^2\rangle_\alpha+ \Zt\langle r_C^2\rangle_t\right)\Sigma'_{q}(-\Blisev)\right].
\label{eq:Liseven_charge_formfactor}
\end{align}
Next we can compute the charge radius of \Liseven using the following definition
\begin{equation}
    \langle r_C^2\rangle_{\Liseven}  = -\frac{3}{ \ZLisev}\frac{\dd^2}{\dd q^2} F^{(\Liseven)}_{\rm C0}(q)\bigg|_{q=0}.
\end{equation}
The second derivative of $\Gamma^{(a)}_{\rm C}$ with respect to $q^2$  at $q=0$ is
\begin{equation}
\begin{aligned}
\frac{\dd^2\Gamma^{(a)}_{\rm C}}{\dd q^2} \bigg|_{q=0}&\sim\frac{\dd^2}{\dd q^2}\int_0^\infty \dd rj_0\left(fqr\right)\left|W_{-\eta_{q}, \frac{3}{2}}(2\gamma_q r)\right|^2\bigg|_{q=0}  \\
 &=  -\frac{f^2}{3}\int_0^\infty \dd r r^2 |W_{-\eta_{q}, \frac{3}{2}}(2\gamma_q r)|^2 = -\frac{f^2}{3}\xi_1,
\end{aligned}
\label{eq:overlap_integral_I}
\end{equation}
where $\xi_1$ is a finite constant that can be solved numerically, yielding $\xi_1 = 7.68585$ fm$^3$. The N$^2$LO charge radius of \Liseven is
\begin{equation}
\begin{aligned}
  \langle r_C^2\rangle_\text{\Liseven} 
  &= \frac{1}{6}\left[\Cpq\right]^2 \frac{\Za f^2+\Zt (1-f)^2}{\ZLisev}\xi_1 + \frac{\calZBlisevq \Xi_q}{\ZLisev} + \calZBlisevq \Sigma'_{q}(-\Blisev)\left(\frac{\Za}{\ZLisev}\langle r_C^2\rangle_\alpha + \frac{\Zt}{\ZLisev}\langle r_C^2\rangle_{\triton} \right).
\end{aligned}
\label{eq:Lithiumseven_chargeradius}
\end{equation}
Relativistic corrections to the charge radius are suppressed by $(\gamma_q/m_R)^2\approx3\times10^{-3}$. Numerically this is smaller than $Q^4$, so these corrections enter beyond the order considered here. The Darwin–Foldy term of the nucleus contributes $3/(4\msev^2)\approx 7\times 10^{-4}$~fm$^2$ to $\langle r_C^2\rangle_{\Liseven}$, or the relative contribution of about $10^{-4}\%$. The Darwin–Foldy and spin-orbit contributions of the constituents are included in their measured charge radii. We therefore neglect all of these corrections.

The expression in \cref{eq:Lithiumseven_chargeradius} contains two unknowns, the ANC $\Cpq$ and the
short-range coefficient $\Xi_q$, so the measured charge radius alone does not
determine either. We therefore impose a second condition. For a two-body system
the charge radius decomposes into the charge-weighted mean-square intercluster
separation and the charge-weighted intrinsic radii of the constituents,
\begin{equation}
 \langle r_C^2\rangle_\text{\Liseven} = \frac{\Za f^2+\Zt (1-f)^2}{\ZLisev}\langle R^2\rangle_\text{\Liseven}+\frac{\Za}{\ZLisev}\langle r_C^2\rangle_\alpha + \frac{\Zt}{\ZLisev}\langle r_C^2\rangle_{\triton} , 
\end{equation}
which is a statement of geometry and folding rather than a dynamical model, and
holds to the order at which we work~\cite{Buck:1977xyz,Mason:2008ka}. In the above equation, $R$ is the inter-cluster distance.
Requiring
\cref{eq:Lithiumseven_chargeradius} to reproduce this decomposition fixes
\begin{equation}
 w_q=\nu_q \left[\Za\langle r_C^2\rangle_\alpha + \Zt\langle r_C^2\rangle_{\triton} \right].  
\end{equation}
This amounts to assuming that the $\alpha$ and the triton retain their free-space
charge distributions inside \Liseven, so that no short-distance contribution to
the charge radius arises beyond their intrinsic sizes at this order. This is a
choice for an N$^2$LO coefficient, not a consequence of the power
counting. 

Using the charge radii of \Liseven, $\alpha$, and triton \cite{ANGELI201369}, we find $\Cpq=3.30\pm(0.10)_{\rm exp}$ \unit{fm^{-1/2}}, where the experimental error results from the propagation of uncertainties in the input parameters. We will delay provide the theoretical error due to omitting N$^3$LO terms in the EFT power counting because it requires a value for the breakdown scale, which we try to quantify from experimental data of the charge form factor in~\cref{sec:liseven_breakdown}. 

To close this section, we obtain a  relationship between ANC and the mean-square intercluster distance $\langle R^2\rangle_{\Liseven}$ at LO,
\begin{equation}
    \langle R^2\rangle_{\Liseven}=\frac{1}{6}\left[\Cpq\right]^2\xi_1,
    \label{eq:Li_rms_intercluster}
\end{equation}
which gives $ \sqrt{\langle R^2\rangle_{\Liseven}}\approx3.73$~fm. This value exceeds $(\langle r_C^2\rangle_\alpha+\langle r_C^2\rangle_{t})^{1/2}=2.43$~fm by a factor of about 1.5. Thus,  the charge distributions of the two clusters overlap only partially.



\subsection{\Liseven's electric quadrupole form factor \label{sec:liseven_quadrupole}}

The three-point vertex functions are already calculated in \cref{eq:three_point_E_form_factor_Lisev}. We obtained the quadrupole form factor of \Liseven up to and including N$^2$LO contributions
\begin{align}
    \frac{1}{2(\msev)^2}F^{(\Liseven)}_{\rm C2}(q) &= -\frac{1}{2}\left[\Cpq\right]^2 \Za \left(1-\frac{\langle r_C^2\rangle_\alpha}{6}q^2\right)\int_0^\infty \dd r ~\frac{j_2(fqr) }{q^2} \left|W_{-\eta_{q},\frac{3}{2}}(2\gamma_{q}r)\right|^2\nonumber\\
    &\quad-\frac{1}{2}\left[\Cpq\right]^2 \Zt \left(1-\frac{\langle r_C^2\rangle_t}{6}q^2\right) \int_0^\infty \dd r ~\frac{j_2\left[(1-f)qr\right] }{q^2} \left|W_{-\eta_{q},\frac{3}{2}}(2\gamma_{q}r)\right|^2.
\label{eq:Liseven_quadrupole_formfactor}
\end{align}
In the limit $q\to 0$, $j_2(x)/x^2$ tends to $1/15$, we obtain
\begin{align}
    \frac{1}{2(\msev)^2}F^{(\Liseven)}_{\rm C2}(0) &= -\left[\Cpq\right]^2  \frac{\Za f^2}{30} \xi_1-\left[\Cpq\right]^2  \frac{\Zt (1-f)^2}{30}\xi_1.
\end{align}
The constituent finite sizes do not contribute at $q=0$, since neither cluster
carries an intrinsic quadrupole moment, so the above equation does not receive N$^2$LO
correction from that source. The result is an LO one. Using the normalization $ \lim_{q\to 0} F^{(\Liseven)}_{\rm C2}(q)/{(\msev)^2} = \quadlisev$, where $\quadlisev$ is the static electric quadrupole moment of \Liseven, we can calculate the static electric quadrupole moment of \Liseven nuclei at LO,
\begin{equation}
    \quadlisev =-\frac{2}{5} \left[\Za f^2 + \Zt (1-f)^2\right]\frac{\left[\Cpq\right]^2}{6}\xi_1=- 3.87\pm (0.24)_{\rm exp}~\unit{fm^2}.
    \label{eq:Lithiumseven_quadrupole}
\end{equation}
We also show that
\begin{equation}
    \quadlisev = -\frac{2}{5} \left[\Za f^2 + \Zt (1-f)^2\right]\langle R^2\rangle_{\text{\Liseven}},
\end{equation}
which is consistent with the formula valid for two-cluster nuclei \cite{buck1985cluster,Mason:2008ka}. The uncertainty in the above prediction is propagated from the experimental
error on $\Cpq$ only. Since $\Cpq$ is already fixed by the charge radius,
this is a parameter-free prediction. The constituent finite-size terms
cancel in the $q\to0$ limit, so the only contribution at N$^2$LO is the
three-point contact operator
$e\,C_Q\,\Pi^\dagger S_i^\dagger S_j \Pi
\left(\nabla_i \nabla_j-\tfrac{\delta_{ij}}{3}\nabla^2\right)A_0$,
whose coefficient $C_Q$ is not fixed by any of the input used here.
Assuming $C_Q$ to be natural, omitting this term induces a relative
uncertainty of order $Q^2$. The predicted value deviates from the
measurement, $\quadlisev = -4.00(3)$~\unit{\fm^2} \cite{STONE20161}, by only
$\sim3\%$, against the $(\gamma_q/\Lambda)^2\sim40\%$ permitted at this order if $\Lambda$ is
identified with the triton breakup scale, $\Lambda\approx142$~MeV. The
coefficient this requires is small for a natural expansion, and remains so
for any assumed order of the first correction.  In the next section we
determine $\Lambda$ explicitly and assign the truncation uncertainties to
the quantities extracted above.

\subsection{The breakdown scale and uncertainty quantification}\label{sec:liseven_breakdown}
We now have calculated both the $F_{\rm C0}$ and $F_{\rm C2}$ form factors, so we
can find the charge form factor measured in elastic electron
scattering. The charge form factor is given by \cite{PhysRev.162.992}
\begin{equation}
    F^2_{\rm ch}(q)=\frac{1}{\ZLisev^2}|F_{\rm C0}(q)|^2+\frac{q^4}{36\ZLisev^2}\left|\frac{F_{\rm C2}(q)}{2\msev^2}\right|^2
\end{equation}
Since both multipoles have been calculated through N$^2$LO, the first omitted
contribution to $F_{\rm ch}^2$ is N$^3$LO and the truncation error is of
relative order $cQ^3$. Without assuming $c\sim1$, we
extract the coefficient required at each measured point for a given $\Lambda$ using 
\begin{equation}
  c_i = \frac{F_{\rm ch,exp}^2(q_i) - F_{\rm ch}^2(q_i)}
             {\big|F_{\rm ch}^2(q_i) - 1\big|\,Q(q_i)^3}.
  \label{eq:ci}
\end{equation}
where $Q = \max\{\gamma_q,\,(1-f)q\}/\Lambda$. The reference scale is the deviation from the point-like value,
since the normalization at $q=0$ is fixed by charge conservation and the
expansion controls only the $q$-dependence. The expansion parameter is the
larger of the binding momentum and the momentum delivered to the triton,
$(1-f)q$, so the uncertainty grows with momentum transfer.

Using the N$^2$LO result of \cref{eq:Liseven_charge_formfactor,eq:Liseven_quadrupole_formfactor} and the data
of Ref.~\cite{PhysRevC.5.391}, and take the root-mean-square over the points with
$q > 0.6~\unit{fm^{-1}}$. We restrict the range because the denominator of
Eq.~(\ref{eq:ci}) vanishes as $q \to 0$, making the ratio numerically unstable at
low momentum transfer, and because the four measurements near $q \simeq 0.27~\unit{fm^{-1}}$ scatter among themselves by considerably more than their quoted uncertainties. For each measured point we solve Eq.~(\ref{eq:ci}) for the coefficient
$c_i$ required to account for the residual at that momentum transfer, and
characterize their size by the root-mean-square value $c^{\rm rms}$ (see \cref{tab:crms})
\setlength{\tabcolsep}{10pt}
\renewcommand{\arraystretch}{1.3}
\begin{table}[h]
\begin{tabular}{cc}\hline\hline
 $\Lambda$ (MeV) & $c^{\rm rms}$ \\ \hline
 142 & 0.28 \\
 180 & 0.58 \\
 200 & 0.80 \\
 215 & 1.00 \\\hline\hline
\end{tabular}
\caption{Root-mean-square expansion coefficient extracted from
Eq.~(\ref{eq:ci}) for different choices of the breakdown scale.}
\label{tab:crms}

\end{table}

With $\Lambda$ identified with the triton breakup momentum, the data
require a coefficient well below unity, which is not what a natural
expansion produces. Since $c_i \propto \Lambda^3$, the scale for which
$c^{\rm rms}=1$ is $\Lambda = 215$~MeV, close to the $\alpha$ excitation
momentum rather than the triton breakup momentum. The
form factor data therefore indicate a breakdown scale close to the $\alpha$
excitation momentum rather than the triton breakup momentum.

This is consistent with the analysis of Ref.~\cite{Poudel:2021mii} for the mirror $\alpha+h$ system.  By fitting LO, NLO, and N$^2$LO results to available elastic scattering cross sections, they could check that the coefficients at successive orders are of similar size and found that 
$\Lambda = 200~\unit{MeV}$ gives a regular convergence pattern for observables. A breakdown scale
exceeding the sub-cluster separation energy indicates that the neglected degrees
of freedom couple only weakly to the elastic channel, which is consistent with
the microscopic calculation of Ref.~\cite{Tao:2025ukl}, where the $\alpha+t$
spectroscopic factor for the \Liseven ground state is found to be close to unity. Therefore, we adopt $\Lambda = 200~\unit{MeV}$ throughout, giving
$Q = \gamma_d/\Lambda\sim 0.44$ at the bound-state pole and an N$^2$LO truncation uncertainty of
$8.5\%$ there, rising with momentum transfer according to Eq.~(\ref{eq:ci}).

Three limitations should be noted. First, because the finite-size operators
first contribute at N$^2$LO with no NLO term, the expansion provides a single
order-to-order step. We can only test whether $c$ is of order unity,
and not whether the coefficients remain of similar size across successive orders,
as was possible in Ref.~\cite{Poudel:2021mii}. Second, the extracted value
depends on the fitted range. For example, including all points with $q > 0.35~\unit{fm^{-1}}$
gives $c^{\rm rms} = 1.22$ rather than $0.80$ at $\Lambda = 200~\unit{MeV}$.
Third, at the lower end of the fitted range the deviations are comparable to the
experimental uncertainties, so the determination is driven by the two highest
momentum transfers, where the truncation error exceeds the measurement error.

We are now in position to provide the theoretical uncertain of extracted observables. First, $\Cpq=3.30\pm(0.10)_{\rm exp} \pm(0.14)_{\rm theo}$ \unit{fm^{-1/2}}. This value is consistent with results obtained from other methods (see Tab.~\ref{tab:Lisev_ANC} for a comparison). 
\setlength{\tabcolsep}{5pt}
\renewcommand{\arraystretch}{1.3}
\begin{table}[h!]
    \centering
    \begin{tabular}{lcccc}\hline\hline
       Method & $\Cpq$ (\unit{fm^{-1/2}}) & $\Cpd$  (\unit{fm^{-1/2}})&$\Cpd/\Cpq$ &Ref. \\\hline
        $\delta$-potential & $3.0 \pm 0.2$ & -- & --&\cite{PhysRevC.100.054307}\\
        R-matrix & $3.49$ &  -- &--&\cite{DESCOUVEMONT2004203}\\
        Variational Monte Carlo & $3.4\pm 0.1$ & $2.65 \pm 0.10$ & 0.779&\cite{Nollett:2001ub}  \\
        NCSMC & 3.49 & 3.16& 0.905&\cite{PhysRevC.100.024304}\\
       Cluster EFT & $3.30\pm (0.17)_{\rm comb}$ & $2.91\pm(0.15)_{\rm comb}$& $0.881$&This work\\\hline
        ${}^3{\rm H}(\alpha,\gamma)\Liseven$ data (model analysis) & $3.57\pm0.15$ & $3.0\pm0.15$&$0.84\pm0.06$&\cite{Igamov:2007svc,IGAMOV2010346}\\\hline\hline
    \end{tabular}
    \caption{Asymptotic normalization coefficients  for $\Liseven (3/2^-) \to t+\alpha$ and $\Liseven (1/2^-)\to t+\alpha$. The unit for the ANC is \unit{fm^{-1/2}}. Uncertainties quoted for this work combine the propagated experimental
uncertainty and the estimated EFT truncation uncertainty in quadrature.  The cluster EFT ratio follows from requiring Eq.~(\ref{eq:BEtwoQ}) with the physical binding energies and carries no independent uncertainty.\label{tab:Lisev_ANC}}
\end{table}

\begin{figure}[h!]
    \includegraphics[width=.47\textwidth]{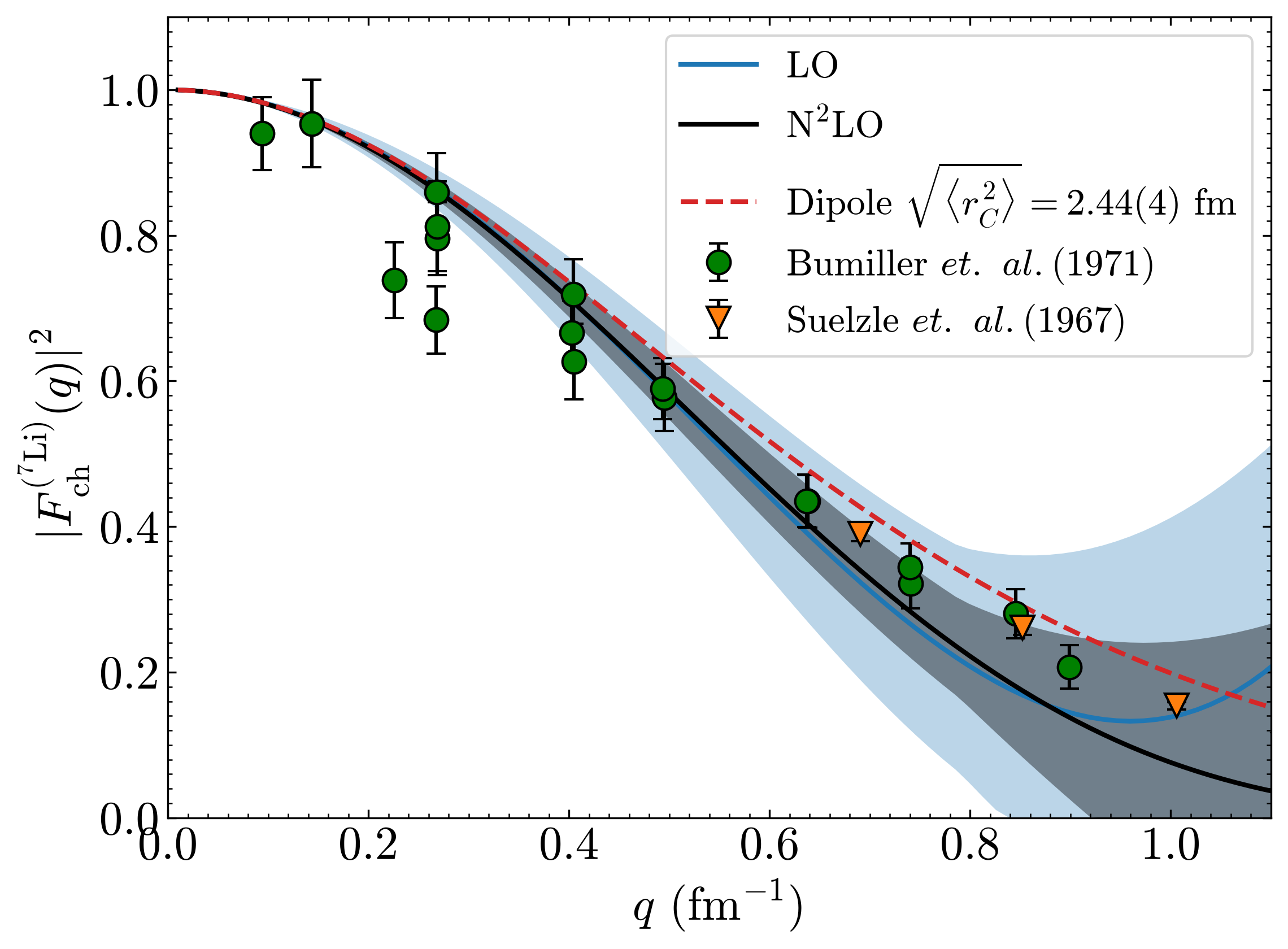}
    \caption{Plot of \Liseven's squared charge form factor (left) and squared magnetic form factor (right) as a function of momentum transfer $q^2$. Experimental data: green circle from Ref.~\cite{PhysRevC.5.391} and orange triangle from Ref.~\cite{PhysRev.162.992}.}
    \label{fig:Liseven_charge_mag}
\end{figure}

Next, we present the calculated charge form factor up to and including N$^2$LO as a function of the momentum transfer in \cref{fig:Liseven_charge_mag}. At low momentum transfer the N$^2$LO curve reproduces the data well and is
indistinguishable from the dipole form obtained from the measured charge
radius alone, as both curves are controlled by
$\langle r_C^2\rangle$ in this region. Beyond $q\approx0.6$~\unit{\per\fm} the two separate,
and the calculation falls systematically below the measurements. The
deviation is nonetheless contained within the estimated truncation
uncertainty at every point, and it is this deviation that the coefficients
of Eq.~(\ref{eq:ci}) quantify.

To close this section, we update our prediction for the static quadrupole moment at LO of
\Liseven\ is $\quadlisev = -3.87 \pm (0.24)_{\rm exp} \pm (0.77)_{\rm theo}
~\unit{fm^2}$, where the theoretical uncertainty is of order 10\%. 

\subsection{\Liseven's magnetic form factor}
\begin{figure}[ht]
\includegraphics[width=.47\linewidth]{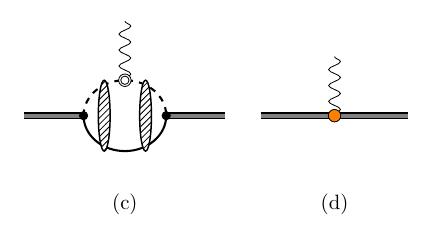}
    \caption{Diagrams for the magnetic form factor $F_{\rm M1}(q)$. For \Liseven, the vertex represented by the double circle corresponds to a vector photon $A_i$ coupling to the triton, which is proportional to $\mut $, while the one depicted by the orange circle corresponds to $\LMone$. Both diagrams appear at LO.  The same diagrams are needed for \Be cluster.}
    \label{fig:M_form_factor_LO}
\end{figure}

At low energies, the magnetic form factor is dominated by the ${\rm M1}$ component. The first contribution arises from the photon coupled to the convective current of the charged $\alpha$ and triton. This coupling comes from gauging the kinetic terms of the $\alpha$ and triton fields with the covariant derivative. It enters through the same one-loop diagrams as the charge form factor (see \cref{fig:E_form_factor}a and \cref{fig:E_form_factor}b). For example, the irreducible three-point function shown in
\cref{fig:E_form_factor}a yields
\begin{equation}
\begin{aligned}
    -i\Gamma^{(a)}_{\rm M} &=ig_q^2 \int\frac{\dd^4k_1\dd^4k_2\dd^4k_3}{(2\pi)^{12}} \left(\vb{k}_3-\frac{(1-f)\vb{q}}{2}\right)_i\left[S^\dagger_{i}S_j\right]_{xy}iS_\alpha\left(E_3, \vb{k}_3\right)iS_t\left(E-E_3,\frac{\vb{q}}{2}-\vb{k}_3\right)\\
    &\quad\times i\chi\left(\vb{k}_3-\frac{f\vb{q}}{2};\vb{k}_2-\frac{f\vb{q}}{2},-\Blisev\right)iS_t(E_2,\vb{k}_2) \left(\vb{k}_1+\frac{(1-f)\vb{q}}{2}\right)_j\\
    &\quad \times \left(\frac{\Za}{m_{\alpha}}\vb{k}_2\right) iS_\alpha\left(E-E_2,-\vb{k}_2+\frac{\vb{q}}{2}\right)iS_\alpha\left(E-E_2,-\vb{k}_2-\frac{\vb{q}}{2}\right)\\
    &\quad\times i\chi\left(\vb{k}_2+\frac{f\vb{q}}{2};\vb{k}_1+\frac{f\vb{q}}{2} ;-\Blisev\right) iS_\alpha\left(E_1, \vb{k}_1\right)iS_t\left(E-E_1,-\frac{\vb{q}}{2}-\vb{k}_1\right).    \end{aligned}
    \label{eq:mag_FF_Lseven}
\end{equation}    
It follows that
\begin{equation}
\begin{aligned}
    -i\Gamma^{(a)}_{\rm M} &=   ig_q^2 \frac{\Za}{\malpha} \int \frac{\dd^3 \vb{k}_1\dd^3 \vb{k}_2\dd^3 \vb{k}_3}{(2\pi)^9} ~\vb{k}_2~\langle \vb{k}_3|G_C(-\Blisev)|\vb{k}_2-f\vb{q}/2\rangle\langle \vb{k}_2+f\vb{q}/2|G_C(-\Blisev)|\vb{k}_1\rangle \left(\vb{k}_3\right)_i\left[S^\dagger_{i}S_j\right]_{xy} \left(\vb{k}_1\right)_j.
\end{aligned}
\label{eq:mag_FF_Lseven_simplified}
\end{equation}
Similar to the previous section, this integral is better solved in the position coordinates
\begin{equation}
\begin{aligned}
    -i\Gamma^{(a)}_{\rm M} =   ig_q^2 \frac{\Za}{\malpha} \int \frac{\dd^3 \vb{k}_1\dd^3 \vb{k}_2\dd^3 \vb{k}_3}{(2\pi)^9}&\dd^3\vb{r}\ \dd^3 \vb{r}'\ \left(\vb{k}_3\right)_i\langle \vb{k}_3|G_C(-\Blisev)|\vb{r})(\vb{r}|\vb{k}_2-f\vb{q}/2\rangle~\vb{k}_2~\\
    &\times \langle \vb{k}_2+f\vb{q}/2|\vb{r}')(\vb{r}'|G_C(-\Blisev)|\vb{k}_1\rangle \left[S^\dagger_{i}S_j\right]_{xy} \left(\vb{k}_1\right)_j.
\end{aligned}
\end{equation}
We arrive at the coordinate-space
integral
\begin{equation}
\begin{aligned}
    -i\Gamma^{(a)}_{\rm M} =   -i \frac{g_q^2 m_r^2\gamma_q^2}{4\pi^2} \Gamma(2+\eta_q)^2 \frac{\Za}{\malpha}\left[S^\dagger_{i}S_j\right]_{xy}\int &\dd^3\vb{r} ~ e^{-if\vb{q}\cdot\vb{r}}\frac{W_{-\eta_q,3/2}(2\gamma_q r)}{r}\vu{r}_i\left(-i\nabla\right) \frac{W_{-\eta_q,3/2}(2\gamma_q r)}{r}  \vu{r}_j.
\end{aligned}
\end{equation}
Projecting onto the magnetic-dipole component (retaining only the part antisymmetric in $i\leftrightarrow j$) and performing the angular integral yield
\begin{equation}
\begin{aligned}
    -i\Gamma^{(a)}_{\rm M} =   -i \frac{g_q^2 m_r^2\gamma_q^2}{2\pi} \Gamma(2+\eta_q)^2 \frac{\Za f}{\malpha }\left[S^\dagger_{i}S_j\right]_{xy}\int_0^\infty &\dd r ~ |W_{-\eta_q,3/2}(2\gamma_q r)|^2 \frac{j_1(fqr)}{fqr} (\delta_{kj}q_i -\delta_{ki}q_j).
\end{aligned}
\end{equation}
It is possible to show that $S^\dagger _i S_j(\delta_{kj}q_i -\delta_{ki}q_j) = \frac{2}{3}i(\vb{J}^{(3/2)}\times\vb{q})^k$, where $J^{(3/2)}_i$’s are the generators of the spin-3/2. Notice that, in the $q\to0$ limit, $j_1(fqr)/fqr$ approaches $1/3$. The above integral is proportional to the divergent energy derivative of the self-energy $\Sigma_q'(-\Blisev)$ that appeared in \cref{eq:divergence_of_threepoint}. The divergence is absorbed by a two-body operator introduced below. Diagram \cref{fig:E_form_factor}b, in which the photon couples to the convection current of the triton, gives the analogous expression with $f\to1-f$ and $\Za/\malpha\to\Zt/\mtthree$. 

At LO, the ${\rm M1}$ amplitude also receives contributions from the magnetic moments of the triton and the photon-auxiliary field vertex, which are described by the following Lagrange density 
\begin{equation}
    \calL_{\rm M1} = \mut t^\dagger_a (\boldsymbol{\sigma}\cdot \vb{B})t_a  - \frac{2}{3}\LMone\Pi^\dagger (\vb{J}^{(3/2)}\cdot \vb{B})\Pi + \LMone'\left[\Pi^\dagger _x\left(S_i^\dagger \right)_{xa}(t_ai\Galilean_i \phi)(\boldsymbol{\sigma}\cdot \vb{B})+\text{h.c.}\right]\,.
\end{equation}
The magnetic field is conventionally defined $\vb{B}=\nabla\times \vb{A}$.  At LO, we have
\begin{equation}
\begin{aligned}
    -i\Gamma^{(c)}_{\rm M} &=ig_q^2\mut  \int\frac{\dd^4k_1\dd^4k_2\dd^4k_3}{(2\pi)^{12}} \left(\vb{k}_3-\frac{f\vb{q}}{2}\right)_i\left[S^\dagger_{i}~i (\boldsymbol{\sigma} \times \vb{q})^kS_j\right]iS_\alpha\left(E_3, \vb{k}_3\right)iS_t\left(E-E_3,\frac{\vb{q}}{2}-\vb{k}_3\right)\\
    &\quad\times i\chi\left(\vb{k}_3-\frac{f\vb{q}}{2};\vb{k}_2-\frac{f\vb{q}}{2},-\Blisev\right)iS_\alpha(E_2,\vb{k}_2) \left(\vb{k}_1+\frac{f\vb{q}}{2}\right)_j\\
    &\quad \times iS_t\left(E-E_2,-\vb{k}_2+\frac{\vb{q}}{2}\right)iS_t\left(E-E_2,-\vb{k}_2-\frac{\vb{q}}{2}\right)\\
    &\quad\times i\chi\left(\vb{k}_2+\frac{f\vb{q}}{2};\vb{k}_1+\frac{f\vb{q}}{2} ;-\Blisev\right) iS_\alpha\left(E_1, \vb{k}_1\right)iS_t\left(E-E_1,-\frac{\vb{q}}{2}-\vb{k}_1\right)\\
     & = i\mut \frac{g_q^2m_r^2\gamma_{q}^2}{4\pi^2}\Gamma\left(2+\eta_{q}\right)^2 \int_0^\infty \dd r~\left[\frac{4\pi}{3}\delta_{ij} j_0\big[(1-f)qr\big] -4\pi j_2\big[(1-f)qr\big]\left(\hat{q}_{i}\hat{q}_{j}-\frac{1}{3}\delta_{ij}\right)\right]\\
     & \qquad \qquad \qquad \qquad\times\left|W_{-\eta_{q},\frac{3}{2}}(2\gamma_{q}r)\right|^2\left[S_i^\dagger(\boldsymbol{\sigma} \times \vb{q})^kS_j\right].
    \end{aligned}
    \label{eq:three_point_M_form_factor}
\end{equation}
Unlike the charge form factor case, the traceless tensor also contributes to the ${\rm M1}$ form factor. Projecting it onto $(\vb{J}^{(3/2)}\times \vb{q})^k$ gives a factor of $-1/5$.  Again, in the limit $q\to0$, $j_0\to1$ and $j_2\to0$, and the remaining integral reduces to $\Sigma_q'(-\Blisev)$. The divergences of $\Gamma^{(a+b)}_{\rm M}$ and $\Gamma^{(c)}_{\rm M}$ are absorbed by the $\LMone$ term (see \cref{fig:M_form_factor_LO}d), which gives the contact vertex 
\begin{equation}
-i\Gamma^{(d)}_{\rm M}=i\LMone[S_i^\dagger S_j]\left(\delta_{ik}q_j-\delta_{jk}q_i\right),
\end{equation}
where we used $S^\dagger _i S_j(\delta_{kj}q_i -\delta_{ki}q_j) = \frac{2}{3}i(\vb{J}^{(3/2)}\times\vb{q})^k$. Thus, the magnetic form factor at LO is given by
\begin{equation}
\begin{aligned}
\frac{e}{2\msev}F^{(\Liseven)}_{\rm M1}(q) =&\ \calZBlisevq\, \frac{g_q^2 m^2_r\gamma_{q}^2}{3\pi}\Gamma\left(2+\eta_{q}\right)^2
\Bigg\{ \mut\int_0^\infty \dd r \left[j_0\big[(1-f)qr\big]-\frac15 j_2\big[(1-f)qr\big]\right]\left|W_{-\eta_{q},\frac32}(2\gamma_{q} r)\right|^2 \\
&+ \frac{\Za m_p f}{\malpha}\int_0^\infty \dd r\, \frac{3j_1(fqr)}{fqr}\left|W_{-\eta_{q},\frac32}(2\gamma_{q} r)\right|^2
\\
&+ \frac{\Zt m_p (1-f)}{\mtthree}\int_0^\infty \dd r\, \frac{3j_1\big[(1-f)qr\big]}{(1-f)qr}\left|W_{-\eta_{q},\frac32}(2\gamma_{q} r)\right|^2 \Bigg\}
\\
&+\calZBlisevq\, \LMone .
\end{aligned}
\end{equation}
In the second and the third lines, we multiplied and divided by the proton mass so that the result is expressed in units of the nuclear magneton $\mu_N$. We can fix the LEC $\LMone$ by the renormalization condition,
\begin{equation}
 \frac{e}{2\msev}F^{(\Liseven)}_{\rm M1}(0)=\mut  + \left[\underbrace{\frac{\Za m_p f}{\malpha }}_{A_{\ell,\alpha}} + \underbrace{\frac{\Zt m_p (1-f)}{\mtthree }}_{A_{\ell,t}}\right],   
\end{equation}
where $\mut =2.97896 \mu_N$ \cite{NevoDinur:2018hdo} and the second piece come form the relative orbital angular momentum $\vb{L}$. Then the magnetic form factor at LO is
\begin{equation}
\begin{aligned}
\frac{ e}{2\msev} F^{(\Liseven,\rm LO)}_{\rm M1}(q)=&~ \frac{1}{6}  \left[\Cpq\right]^2\mut\int_0^\infty \dd r\left\{j_0\big[(1-f)qr\big]-1-\frac{1}{5}j_2\big[(1-f)qr\big]\right\}\left|W_{-\eta_{q},\frac{3}{2}}(2\gamma_{q} r)\right|^2 \\
    &+\frac{1}{6}  \left[\Cpq\right]^2A_{\ell,\alpha}\int_0^\infty \dd r\, \left[\frac{3j_1(fqr)}{fqr}-1\right]\left|W_{-\eta_{q},\frac32}(2\gamma_{q} r)\right|^2
\\
&+ \frac{1}{6}  \left[\Cpq\right]^2A_{\ell,t}\int_0^\infty \dd r\, \left\{\frac{3j_1\big[(1-f)qr\big]}{(1-f)qr}-1\right\}\left|W_{-\eta_{q},\frac32}(2\gamma_{q} r)\right|^2 \\
&+\mut  + A_{\ell,\alpha}+ A_{\ell,t}. 
\end{aligned}
\end{equation}
\begin{figure}[t]
\includegraphics[width=.5\linewidth]{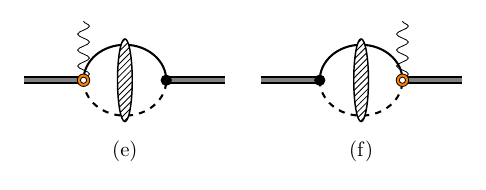}
\caption{Diagrams contributing to the \Liseven magnetic form factor arising from the photon coupled to the dimer-$\alpha t$ vertex $\LMone'$ (orange ring). All diagrams appear at N$^2$LO. }
    \label{fig:M_form_factor_NLO}
\end{figure}
At N$^2$LO, the magnetic form factor receives contributions from diagrams in Figs.~\ref{fig:M_form_factor_NLO}c and \ref{fig:M_form_factor_NLO}d, which are given by
\begin{equation}
\begin{aligned}
    -i\Gamma^{(e+f)}_{\rm M1} &=ig_q \LMone'\int\frac{\dd^4k_1\dd^4k_2}{(2\pi)^{8}} \left(\vb{k}_2-\frac{f\vb{q}}{2}\right)_i\left[S^\dagger_{i}~i (\boldsymbol{\sigma} \times \vb{q})^kS_j\right]iS_t\left(E_2, \vb{k}_2\right)iS_\alpha\left(E-E_2,\frac{\vb{q}}{2}-\vb{k}_2\right)\\
    &\quad\times i\chi\left(\vb{k}_2-\frac{f\vb{q}}{2};\vb{k}_1-\frac{f\vb{q}}{2},-B_\text{\Liseven}\right)\left(\vb{k}_1+\frac{f\vb{q}}{2}\right)_j iS_t\left(E_1, \vb{k}_1\right)iS_\alpha\left(E-E_1,-\frac{\vb{q}}{2}-\vb{k}_1\right)\\
     & = 2i g_q \LMone'\Sigma_q(-\Blisev)\left[ S_i^\dagger(\boldsymbol{\sigma} \times \vb{q})^kS_i\right].
    \end{aligned}
    \label{eq:three_point_M_form_factor_NLO}
\end{equation}
We can fix $\LMone'$ using the \Liseven magnetic dipole moment,
\begin{equation}
    \LMone'= \frac{1}{2g_q \calZBlisevq\Sigma_q(-\Blisev)} \left[\museven  -\left(\mut +A_{\ell,\alpha}+ A_{\ell,t}\right)\right].
\end{equation}
\begin{figure}[h!]
    \includegraphics[width=.5\textwidth]{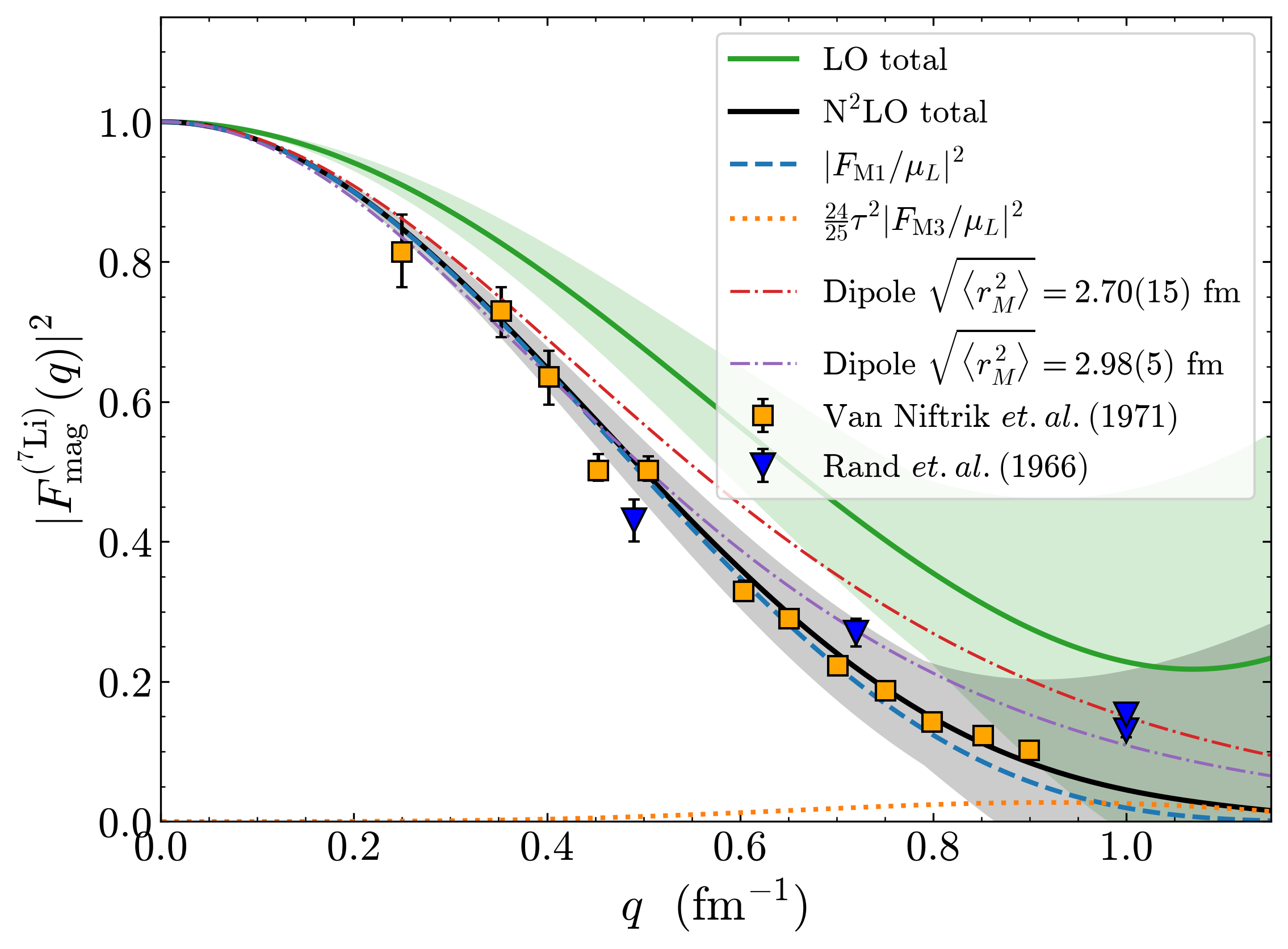}
    \caption{Plot of \Liseven's squared magnetic form factor  as a function of momentum transfer $q$. Experimental data: orange square from Ref.~\cite{van1971magnetization} and blue triangle from Ref.~\cite{Rand:1966zz}. The data of Ref.~\cite{Rand:1966zz} contain the unresolved inelastic excitation of the $0.478$~MeV level. }
    \label{fig:Liseven_mag_only}
\end{figure}
At this order, there is also a contribution from the finite magnetization radius of the triton and the finite charge radii of both $\alpha$ and triton. Similar to the previous section on the charge form factor of \Liseven, 
the magnetic form factor up to and including N$^2$LO terms is
\begin{equation}
\begin{aligned}
\frac{e }{2\msev}F^{(\Liseven,\rm N^2LO)}_{\rm M1}(q) =&~\frac{1}{6} \mut \left[\Cpq\right]^2\int_0^\infty \dd r\left\{j_0\big[(1-f)qr\big]-\frac{1}{5}j_2\big[(1-f)qr\big]-1\right\}\left|W_{-\eta_{q},\frac{3}{2}}(2\gamma_{q} r)\right|^2 \left(1-\frac{\langle r_M^2\rangle_t}{6}q^2\right) \\
&+\frac{1}{6}A_{\ell,\alpha} \left[\Cpq\right]^2\int_0^\infty \dd r\, \left[\frac{3j_1(fqr)}{fqr}-1\right]\left|W_{-\eta_{q},\frac32}(2\gamma_{q} r)\right|^2\left(1-\frac{\langle r_C^2\rangle_\alpha}{6}q^2\right)
\\
&+ \frac{1}{6}A_{\ell,t} \left[\Cpq\right]^2\int_0^\infty \dd r\, \left\{\frac{3j_1\big[(1-f)qr\big]}{(1-f)qr}-1\right\}\left|W_{-\eta_{q},\frac32}(2\gamma_{q} r)\right|^2 \left(1-\frac{\langle r_C^2\rangle_t}{6}q^2\right) \\
&+\museven -\frac{q^2}{6} \bigg[\mut\langle r^2_M\rangle_t +A_{\ell,\alpha}\langle r_C^2\rangle_\alpha+A_{\ell,t}\langle r_C^2\rangle_t \bigg], 
\end{aligned}
\label{eq:liseven_magnetic_formfactor}
\end{equation}
where $\sqrt{\langle r_M^2\rangle_t} =1.840 \pm 0.181$ fm is the rms magnetic radius of triton \cite{AMROUN1994596}. It is straightforward to show that the magnetization radius of \Liseven at LO and  N$^2$LO are respectively
\begin{equation}
 \langle r_M^2\rangle_{\rm LO} =\frac{\left[\Cpq\right]^2}{\mut+A_{\ell,\alpha}+ A_{\ell,t}}\left[\frac{9}{50}\mut\left(1-f\right)^2+\frac{1}{10} A_{\ell,\alpha}f^2+ \frac{1}{10}A_{\ell,t}(1-f)^2\right]\xi_1,
\end{equation}
and 
\begin{equation}
 \langle r_M^2\rangle_{\rm N^2LO}  = \frac{\left[\Cpq\right]^2 }{\museven}\left[\frac{9}{50}\mut\left(1-f\right)^2+\frac{1}{10} A_{\ell,\alpha}f^2+ \frac{1}{10}A_{\ell,t}(1-f)^2\right]\xi_1 + \frac{\mut }{\museven}\langle r^2_M\rangle_t+\frac{A_{\ell,\alpha} }{\museven}\langle r^2_C\rangle_\alpha+\frac{A_{\ell,t} }{\museven}\langle r^2_C\rangle_t.  
 \label{eq:liseven_magnetic_radius}
\end{equation}
We obtain $\sqrt{\langle r_M^2\rangle_{\Liseven}}= 2.86 \pm(0.13)_{\rm exp}\pm(0.12)_{\rm theo}$ fm (N$^2$LO) (see Tab.~\ref{tab:liseven_magnetic_radius} for a comparison to previous studies). We see that there is a mild tension between experimental data there is no consensus yet about this quantity, which differ by about $2\sigma$. Our theoretical calculation is consistent with both the extractions of Refs.~\cite{Rand:1966zz} and \cite{van1971magnetization} within $1\sigma$.  The present uncertainty, which results from the triton magnetization radius and the EFT truncation, is not yet sufficient to discriminate between them.

\setlength{\tabcolsep}{5pt}
\renewcommand{\arraystretch}{1.3}
\begin{table}[h]
    \begin{tabular}{l|l|c|c}\hline\hline
    &\multicolumn{1}{c|}{Method}    &  $\langle r_M^2\rangle^{1/2}$ (fm) & Refs.\\\hline
   \multirow{3}{*}{Theory} &$\alpha + t$  cluster model & $3.04$ &\cite{kajino1984electromagnetic}\\
    &Cluster model& $2.78\pm0.13$ & \cite{buck1985cluster}\\
    & Cluster EFT (N$^2$LO)  & $2.86 \pm(0.18)_{\rm comb}$  &This work\\
    \hline
    \multirow{3}{*}{Exp.}& & $2.70\pm0.15$ & \cite{Rand:1966zz}\\
    & Elastic $e$-\Liseven scattering & $2.69\pm0.13$ &\cite{Rand:1966zz}\\
    & & $2.98\pm0.05$ &\cite{van1971magnetization}\\\hline\hline
    \end{tabular}
    \caption{Summary of previous studies of \Liseven rms magnetization radius. Uncertainties quoted for this work combine the propagated experimental
uncertainty and the estimated EFT truncation uncertainty in quadrature.}
    \label{tab:liseven_magnetic_radius}
\end{table}
Lastly, \cref{eq:three_point_M_form_factor} also gives a parameter-free contribution to the octupole form factor $F^{(\Liseven)}_{\rm M3}$ at LO. In addition, a portion of the N$^2$LO contribution to $F_{\rm M3}$ comes from the triton magnetization form factor. 
\begin{equation}
    F^{(\Liseven)}_{\rm M3}(q)=-\frac{\mut}{q^2}\left[\Cpq\right]^2\int_0^\infty \dd r j_2\big[(1-f)qr\big]\left|W_{-\eta_q,\frac{3}{2}}(2\gamma_q r)\right|^2\left(1-\frac{\langle r_M^2\rangle_t}{6}q^2\right).
\end{equation}
We can obtain the octupole moment at LO via
\begin{equation}
    \Omega_L=-\frac{3}{2\msev^2}F^{(\Liseven)}_{\rm M3}(0)=\frac{1}{10}\mu_T [\Cpq]^2(1-f)^2\xi_1=8.2\pm(0.5)_{\rm exp}\pm(1.6)_{\rm th} ~\mu_N{\rm fm}^2.
\end{equation}
This gives the ratio $\Omega_L/\museven=2.51\pm(0.15)_{\rm exp}\pm(0.50)_{\rm th}$~\unit{fm^2} for \Liseven, in agreement with the experimental value of $2.8\pm0.5$ extracted in \cite{Rand:1966zz}. 
The remaining N$^2$LO piece is an octupole contact term, however there is no data to constrain it. The omission of this term puts the truncation uncertainty of $F^{(\Liseven)}_{\rm M3}$ at the $Q^2$-level. The magnetic form factor is then given by
\begin{equation}
    |F_{\rm mag}(q)|^2=\left| \frac{F_{\rm M1}(q)}{F_{\rm M1}(0)}\right|^2+\frac{24q^4}{25}\left|\frac{F_{\rm M3}(q)}{4\msev^2F_{\rm M1}(0)}\right|^2.
\end{equation}
\Cref{fig:Liseven_mag_only} presents our calculation of the magnetic form factor square as a function of the momentum transfer. The shift between the two orders is dominated by the triton magnetization radius. At \nnlo, our analytical curve agrees with the elastic data \cite{van1971magnetization} within uncertainties over the whole range. The M3 contribution becomes more noticeable for $q\gtrsim0.8$ \unit{fm^{-1}} as expected.  For comparison, \cref{fig:Liseven_mag_only} also shows dipole form factors, $|F_{\rm M}(q)|^2=(1+q^2\langle r_M^2\rangle/12)^{-4}$, evaluated with the two experimental magnetic radii, $2.70$~fm~\cite{Rand:1966zz} and $2.98$~fm~\cite{van1971magnetization}. Our result coincides with both curves at low momentum regime.

\subsection{The reduced Coulomb transition probability \ensuremath{B({\rm C2}): 3/2^-\to 1/2^-}\label{sec:liseven_BE2}}

The ground state of \Liseven can be excited via the absorption of electromagnetic radiation. Such transitions are sensitive to the internal structure of the nucleus. Therefore, they can provide a crucial benchmark for theoretical calculations. The electric quadrupole transition of \Liseven to its first excited state has already been measured in several experiments. Experimentally, the momentum-dependent $B({\rm C2},q)$ is obtained from electron-nucleus scattering. In contrast, the reduced electric transition strength, $B({\rm E2})$, is defined at the real-photon point. Because the virtual and real photon regimes are connected by Siegert theorem \cite{Siegert:1937yt} in the long-wavelength limit, extrapolating the finite-$q$ electron scattering measurements to  ($q\to 0)$ recovers the $B({\rm E2})$ value. The recommended value $B({\rm E2})= 8.3 \pm 0.5~\unit{e^2fm^4}$ obtained from TUNL Nuclear Data Evaluation Group \cite{TILLEY20023}. Later Coulomb-excitation analyses, however, gave a smaller value $B({\rm E2}) = 7.96 \pm 0.12~\unit{e^2fm^4}$ ~\cite{VERMEER1989212}.




\begin{figure}[h]
    \centering
\includegraphics[width=0.6\linewidth]{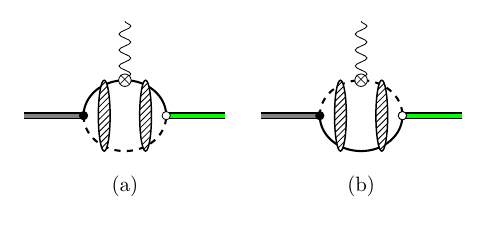}
    \caption{Diagrams for the \Liseven electric quadrupole excitation from \Liseven ground state to the its first excited state. The green double line indicates the dressed propagator of the $\Xi$ field and the white circle represents the insertion of the $g_d$ vertex. All diagrams are of order $\calO(Q)$.}
    \label{fig:Btwo_transition}
\end{figure}

The loop-diagrams consist of a $\alpha$-$t$ bubble, where the external photon line couples to either the $\alpha$ or the triton. In momentum-space the loop-diagram is given by
\begin{equation}
\begin{aligned}
    -i\Gamma^{(a)}_{\rm C2} &=ig_qg_d\Za \int\frac{\dd^4k_1\dd^4k_2\dd^4k_3}{(2\pi)^{12}} \left(\vb{k}_3-\frac{f\vb{q}}{2}\right)_i\left[S^\dagger_{i}\sigma_j\right]_{xa}iS_\alpha\left(E_3, \vb{k}_3\right)iS_t\left(E-E_3,\frac{\vb{q}}{2}-\vb{k}_3\right)\\
    &\quad\times i\chi\left(\vb{k}_3-\frac{f\vb{q}}{2};\vb{k}_2-\frac{f\vb{q}}{2},-\Blisev\right)iS_t(E_2,\vb{k}_2) \left(\vb{k}_1+\frac{f\vb{q}}{2}\right)_j\\
    &\quad \times iS_\alpha\left(E-E_2,-\vb{k}_2+\frac{\vb{q}}{2}\right)iS_\alpha\left(E-E_2,-\vb{k}_2-\frac{\vb{q}}{2}\right)\\
    &\quad\times i\chi\left(\vb{k}_2+\frac{f\vb{q}}{2};\vb{k}_1+\frac{f\vb{q}}{2} ;-\Blisev\right) iS_\alpha\left(E_1, \vb{k}_1\right)iS_t\left(E-E_1,-\frac{\vb{q}}{2}-\vb{k}_1\right)\\
     &= -i\Za\frac{g_dg_q m_r^2}{\pi}\gamma_{q}\gamma_{d}\Gamma\left(2+\eta_{q}\right)\Gamma\left(2+\eta_{d}\right)\left[S^\dagger_{j}\sigma_{i}\right]_{xa}\hat{q}_{j}\hat{q}_{i} \int_0^\infty \dd r j_2(fqr) W_{-\eta_{q},\frac{3}{2}}(2\gamma_{q} r)W_{-\eta_{d},\frac{3}{2}}(2\gamma_{d} r).
    \end{aligned}
\end{equation}
The above integral is finite and can be calculated numerically. Similarly, 
\begin{equation}
\begin{aligned}
-i\Gamma^{(b)}_{\rm C2}  &= -i\Zt\frac{g_dg_q m_r^2}{\pi} \gamma_{q}\gamma_{d}\Gamma\left(2+\eta_{q}\right)\Gamma\left(2+\eta_{d}\right)\left[S^\dagger_{j}\sigma_{i}\right]_{ba}\hat{q}_{j}\hat{q}_{i}\int_0^\infty \dd r j_2[(1-f)qr] W_{-\eta_{q},\frac{3}{2}}(2\gamma_{q}r)W_{-\eta_{d},\frac{3}{2}}(2\gamma_{d} r).
\end{aligned}
\end{equation}
The form factor calculated from the EFT at N$^2$LO is 
\begin{equation}
\begin{aligned}
\frac{F'^{(\Liseven)}_{\rm C2}(q)}{2(\msev)^2} =&~ \frac{1}{q^2}\sqrt{\calZBlisevd}\left[\Gamma^{(a)}_{\rm C2} +\Gamma^{(b)}_{\rm C2} \right]\sqrt{\calZBlisevq}\\
=&~  \frac{1}{2}\Cpq\Cpd \Za \left(1-\frac{\langle r_C^2\rangle_\alpha}{6}q^2\right)\int_0^\infty \dd r \frac{j_2(fqr)}{q^2} W_{-\eta_{q},\frac{3}{2}}(2\gamma_{q} r)W_{-\eta_{d},\frac{3}{2}}(2\gamma_{d} r)    \\
&+\frac{1}{2}\Cpq\Cpd\Zt \left(1-\frac{\langle r_C^2\rangle_t}{6}q^2\right)\int_0^\infty \dd r \frac{j_2[(1-f)qr]}{q^2} W_{-\eta_{q},\frac{3}{2}}(2\gamma_{q} r)W_{-\eta_{d},\frac{3}{2}}(2\gamma_{d} r). 
\end{aligned}
\label{eq:Liseven_quadrupoletransition}
\end{equation}
The inelastic ${\rm C2}$ form factor is then given by (see Ref.~\cite{van1971magnetization} for an example)
\begin{equation}
    \left[\frac{F'^{(\Liseven)}_{\rm C2}(q)}{(\msev)^2}\right]^2= \frac{16\pi}{25}B\left({\rm C2},q\right).
\end{equation}
In the limit $q\to 0$ the reduced transition probability at LO becomes
\begin{equation}
    B({\rm E2})=\lim_{q\to 0} B({\rm C2},q) =\frac{25}{16\pi}\left[\Cpq \Cpd \frac{\Zt (1-f)^2+\Za f^2}{15}\xi_1'\right]^2= \frac{25}{16\pi} \left(\frac{\Cpd}{\Cpq}\right)^2\left(\frac{\xi_1'}{\xi_1}\right)^2 \left[\quadlisev\right]^2,
     \label{eq:BEtwoQrelation}
\end{equation}
where the $\xi_1'$ is a finite integral, $\xi_1' = \int_0^\infty \dd r~r^2 W_{-\eta_d,\frac{3}{2}}(2\gamma_{d} r) W_{-\eta_{q},\frac{3}{2}}(2\gamma_{q} r)\approx8.72816~\text{fm}^3$. The constituent finite-size
corrections vanishes at $q=0$ and therefore do not contribute to $B(\rm E2)$. The reduced transition probability and the static quadrupole
moment are not independent at leading order. In the degenerate limit $\gamma_d=\gamma_q$ and $\Cpd=\Cpq$, this reduces to (see Ref.~\cite{van1971magnetization} for an example),
\begin{equation}
 B({\rm E2}) \simeq \frac{25}{16\pi}\left[\quadlisev\right]^2 .
  \label{eq:BEtwoQ}
\end{equation}

Taking $B({\rm E2})=8.3\pm0.5$ \unit{e^2fm^4} and $\quadlisev=-4.00\pm0.03$ \unit{fm^2} as input, \cref{eq:BEtwoQrelation} gives $(\Cpd\xi_1')^2/(\Cpq\xi_1)^2=1.04\pm0.06$, implying that the relation (\ref{eq:BEtwoQ}) holds well even though the spin-orbit splitting is nonzero. Therefore we assume that \cref{eq:BEtwoQ} continues to hold for the physical binding energies of both states. This determines the ratio of asymptotic
normalization coefficients,
\begin{equation}
  \frac{\Cpd}{\Cpq} = \frac{\xi_1}{\xi_1'}=0.881.
\end{equation}
This value carries no independent uncertainty. Their reliability is set by how well that relation holds. For comparison, the NCSMC
calculation of Ref.~\cite{PhysRevC.100.024304} gives $0.905$, the variational Monte Carlo result of
Ref.~\cite{Nollett:2001ub} gives $0.78$, and the phase-shift extraction of Ref.~\cite{Igamov:2009eh} gives
$0.84$. Combining the ratio with $\Cpq$ yields the excited-state coefficient,
\begin{equation}
  \Cpd = 2.91 \pm (0.09)_{\rm exp} \pm (0.12)_{\rm theo}~\unit{fm^{-1/2}} ,
\end{equation}
where the theoretical uncertainty is inherited from $\Cpq$ and is therefore
fully correlated with it.

\begin{figure}[h]
    \centering
    \includegraphics[width=.47\linewidth]{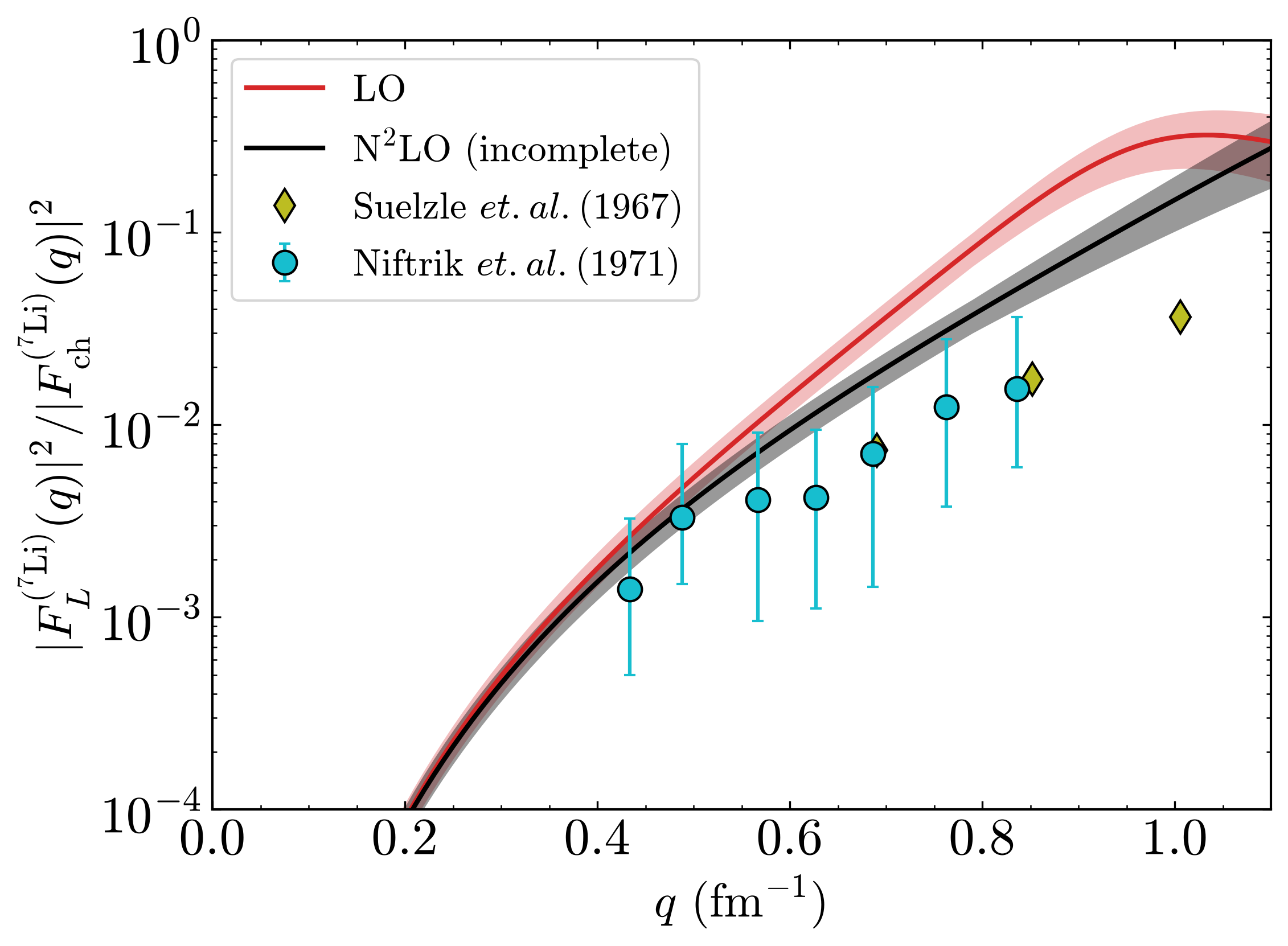}
    \caption{The plot of the ratio of the longitudinal $\rm(C2')$ form factor 
    for excitation of the 0.478 MeV level to the elastic charge form factor as a function of momentum transfer.   The cyan circle data is taken from Ref.~\cite{van1971magnetization} and the olive diamond data is taken from Ref.~\cite{PhysRev.162.992}. \nnlo (partial) includes the finite-size corrections of the constituents but not the contact \nnlo term. Its band reflects an $\mathcal{O}(Q^2)$ truncation uncertainty.}
    \label{fig:longitudinal_ratio}
\end{figure}

Finally, the longitudinal form factor for excitation of the
$478~\unit{keV}$ level is given by \cite{van1971magnetization}
\begin{equation}
    \left|F^{(\Liseven)}_L(q)\right|^2 = \frac{q^4}{36}\left[\frac{F'^{(\Liseven)}_{\rm C2}(q)}{\ZLisev(\msev)^2} \right]^2,
\end{equation}
The LO result is shown in
\cref{fig:longitudinal_ratio}, providing a momentum-dependent test of the same
matrix element. Since the \nnlo contact term is not included, the longitudinal form factor is incomplete at \nnlo, and we assign it a truncation uncertainty of order $Q^2$.

\section{Beryllium-7 \label{sec:beseven}}
The \Be is another interesting stable nucleus with an atomic number of 7. The $S$-factor for the reaction $\Hethree(\alpha,\gamma)\Be$ at astrophysical energies is critical for understanding the solar neutrino flux detected on Earth.  $\Be$ has been widely considered as an $\alpha+\Hethree$ system \cite{kajino1984electromagnetic, PhysRevC.100.054307,Higa:2016igc,Igamov:2009eh}.  Similar to \Liseven, \Be's ground state has quantum numbers $J^\pi = 3/2^-$ and is bound by $\Bbe=1.5866$ MeV with respect to the $\alpha+\Hethree$ threshold \cite{TILLEY20023}. Its first excited state has quantum numbers $J^\pi = 1/2^-$ and is bound by $\Bbe^*=1.1575$ MeV with respect to the threshold \cite{TILLEY20023}. Both energies are considerably smaller than the proton separation energy in \Hethree ($S_p=5.5$ MeV). However, as discussed above and supported by Ref.~\cite{Poudel:2021mii}, whose analysis concerns the $\alpha+h$ system itself, we will use the same breakdown scale $\Lambda=200$~MeV for \Be. With the binding momentum $\gamma_q\simeq71$~MeV, this gives $Q\simeq0.36$.
The relative orbital angular momentum of $\alpha$ and \Hethree for both states is $\ell=1$. As the calculations of \Be's EM form factors here replicate the procedure in Sec.~\ref{sec:lithium_seven}, only key results will be highlighted. First, the charge radius of \Be at N$^2$LO can be obtained similarly to \cref{eq:Lithiumseven_chargeradius}, which reads
\begin{equation}
 \langle r_C^2\rangle_{\Be} = \frac{\Za (f^*)^2+\Zhe (1-f^*)^2}{\ZBe}\frac{1}{6}\left[\Cpq'\right]^2 \xi_2+\frac{\Za}{\ZBe}\langle r_C^2\rangle_\alpha + \frac{\Zhe}{\ZBe}\langle r_C^2\rangle_{\Hethree} , 
\end{equation}
where $f^*=\mhthree /\mbe$, $\Cpq'$ is the ANC for the $\Be\to\alpha+\Hethree$ process, and $ \xi_2 = 6.06559$ \unit{fm^3} is the value of the integral similar to the one given in \cref{eq:overlap_integral_I}. Given the experimental measurements of the \Be charge radius $\sqrt{\langle r_C^2\rangle_{\Be}} =2.647(17)$ fm \cite{Nortershauser:2008vp} and the \Hethree charge radius $\sqrt{\langle r_C^2\rangle_{\Hethree}}=1.9661(30)$ fm \cite{ANGELI201369}, we find that
\begin{equation}
    \Cpq'=3.77\pm (0.05)_{\rm exp}\pm (0.09)_{\rm theo}~\text{fm}^{-1/2}.
\end{equation}
We compare our result with other approaches for the ground state ANC and summarize in Tab.~\ref{tab:BeANC}. 
\begin{table}[h!]
    \centering
    \begin{tabular}{lcccc}\hline\hline
       Method & $\Cpq'$ (\unit{fm^{-1/2}}) & $\Cpd'$ (\unit{fm^{-1/2}}) &$\Cpd'/\Cpq'$ &Ref. \\\hline
        $\delta$-potential & $3.6 \pm 0.1$ & -- & --&\cite{PhysRevC.100.054307}\\
        R-matrix & $3.79$ &  $3.79$&--&\cite{DESCOUVEMONT2004203}\\
        Variational Monte Carlo & $3.55\pm0.15 $ & $2.90\pm0.10$ & $0.82\pm0.04$&\cite{Nollett:2001ub}  \\
        NCSMC & $3.91$ & $3.53$& $0.903$&\cite{Atkinson:2024zrm}\\
       Cluster EFT & $3.77\pm(0.10)_{\rm comb}$ & $3.41\pm(0.09)_{\rm comb}$&$0.904$ &This work\\\hline
        ${}^3{\rm He}(\alpha,\gamma)\Be$ data (model analysis) 
          & $4.83^{+0.1}_{-0.26} $ & $3.99^{+0.08}_{-0.19}$ & 0.826&\cite{PhysRevC.85.045807}  \\\hline\hline
    \end{tabular}
    \caption{Asymptotic normalization coefficients  for $\Be (3/2^-) \to \Hethree+\alpha$ and $\Be (1/2^-)\to \Hethree+\alpha$. The R-matrix analysis assumes equal ANCs for both states. The cluster EFT ratio follows from Eq.~(\ref{eq:BEtwoQ}) and carries no independent uncertainty.}
    \label{tab:BeANC}
\end{table}

The root-mean-square intercluster distance $\langle R^2\rangle_{\Be}$ at LO (see~\cref{eq:Li_rms_intercluster}) is about 3.81 fm. 

Similar to \cref{eq:liseven_magnetic_formfactor}, the \Be's magnetic form factor up to and including N$^2$LO is given by
\begin{equation}
\begin{aligned}
\frac{e }{2\msev}F^{(\Be,\rm N^2LO)}_{\rm M1}(q) =&~\frac{1}{6} \muh \left[\Cpq\right]^2\int_0^\infty \dd r\left[j_0\big[(1-f^*)qr\big]-\frac{1}{5}j_2\big[(1-f^*)qr\big]-1\right]\left|W_{-\eta_{q},\frac{3}{2}}(2\gamma_{q} r)\right|^2 \left(1-\frac{\langle r_M^2\rangle_h}{6}q^2\right) \\
&+\frac{1}{6}A_{\ell,\alpha} \left[\Cpq\right]^2\int_0^\infty \dd r\, \left[\frac{3j_1(f^*qr)}{f^*qr}-1\right]\left|W_{-\eta_{q},\frac32}(2\gamma_{q} r)\right|^2\left(1-\frac{\langle r_C^2\rangle_\alpha}{6}q^2\right)
\\
&+ \frac{1}{6}A_{\ell,h} \left[\Cpq\right]^2\int_0^\infty \dd r\, \left[\frac{3j_1\big[(1-f^*)qr\big]}{(1-f^*)qr}-1\right]\left|W_{-\eta_{q},\frac32}(2\gamma_{q} r)\right|^2 \left(1-\frac{\langle r_C^2\rangle_h}{6}q^2\right) \\
&+\mub -\frac{q^2}{6} \left[\muh\langle r^2_M\rangle_h +A_{\ell,\alpha}\langle r_C^2\rangle_\alpha+A_{\ell,h}\langle r_C^2\rangle_h \right], 
\end{aligned}
\label{eq:beseven_magnetic_formfactor}
\end{equation}
In the above expression, we have already used the magnetic dipole moments of \Be and \Hethree, which are $\mub=-1.3995(5)\mu_N$ \cite{Nortershauser:2008vp}  and $\muh=-2.127\mu_N$ \cite{NevoDinur:2018hdo}, respectively, to fix the undetermined parameters. The calculated magnetization radius of \Be at N$^2$LO is
\begin{equation}
 \langle r_M^2\rangle_{\Be} = \frac{\left[\Cpq'\right]^2 }{\mub}\left[\frac{9}{50}\muh\left(1-f^*\right)^2+\frac{1}{10} A_{\ell,\alpha}f^2+ \frac{1}{10}A_{\ell,h}(1-f^*)^2\right]\xi_2 + \frac{\muh }{\mub}\langle r^2_M\rangle_h+\frac{A_{\ell,\alpha} }{\mub}\langle r^2_C\rangle_\alpha+\frac{A_{\ell,h} }{\mub}\langle r^2_C\rangle_h.  
 \label{eq:Be_magnetic_radius}
\end{equation}
Using $\sqrt{\langle r^2_M\rangle_{\Hethree}}=1.976(47)$ fm \cite{Sick:2014yha}, we obtain $\sqrt{\langle r_M^2\rangle_{\Be}}= 3.34\pm(0.05)_{\rm exp}\pm(0.08)_{\rm theo.}$ fm. We compare the resulting magnetic form factor to an analysis from quantum Monte Carlo calculations using the Norfolk NV2+3-Ia$^*$ interaction and $\chi$EFT electromagnetic currents up to N$^3$LO \cite{Chambers-Wall:2024fha, Chambers-Wall:2024uhq}\footnote{Our result is related to Ref.~\cite{Chambers-Wall:2024fha}'s  definition of the magnetic form factor via $|F^*_{\rm mag}(q)|^2 = \frac{5}{18\pi}\left[\left(\frac{q}{2m_N}\right)\frac{ eF^{(\Be)}_{\rm mag}(q)}{2\mbe \mu_N}\right]^2$.}.

The \Be quadrupole moment still has not been measured, so it is interesting to calculate this quantity using cluster EFT. In analogy to \cref{eq:Lithiumseven_quadrupole}, we find that
\begin{equation}
    \quadbe =-\frac{2}{5} \left[\Za (f^*)^2 + \Zhe (1-f^*)^2\right]\frac{\left[\Cpq'\right]^2}{6}\xi_2=- 5.85\pm(0.16)_{\rm exp}\pm(0.59)_{\rm theo}~\unit{fm^2}.
\end{equation}
Let us compare our results with other approaches. The cluster model calculation in Ref.~\cite{Mertelmeier:1986egp} found that $ \quadbe=-5.84~\unit{fm^2}$.  The \textit{ab initio} computation in Ref.~\cite{Dohet-Eraly:2015ooa} and more recently in Ref.~\cite{Atkinson:2024zrm} found $ \quadbe=-6.14~\unit{fm^2}$ and   $ \quadbe=-6.31~\unit{fm^2}$, respectively. In particular, the later calculation is based on the $\alpha$-$\Hethree$-clustering structure of \Be, whose microscopic interactions are built from chiral two- and three-nucleon forces. It also takes into account many more eigenstates of \Be and a compensation for the deformation of cluster constituents during the reaction. Cluster EFT only considered contact interactions, which significantly reduces computational effort, yet it seems to capture over 90\% of the results. This approach can provide clearer insights into the potential sources of relative contributions in \textit{ab initio} calculations. 

\begin{figure}[h!]
    \includegraphics[width=.32\textwidth]{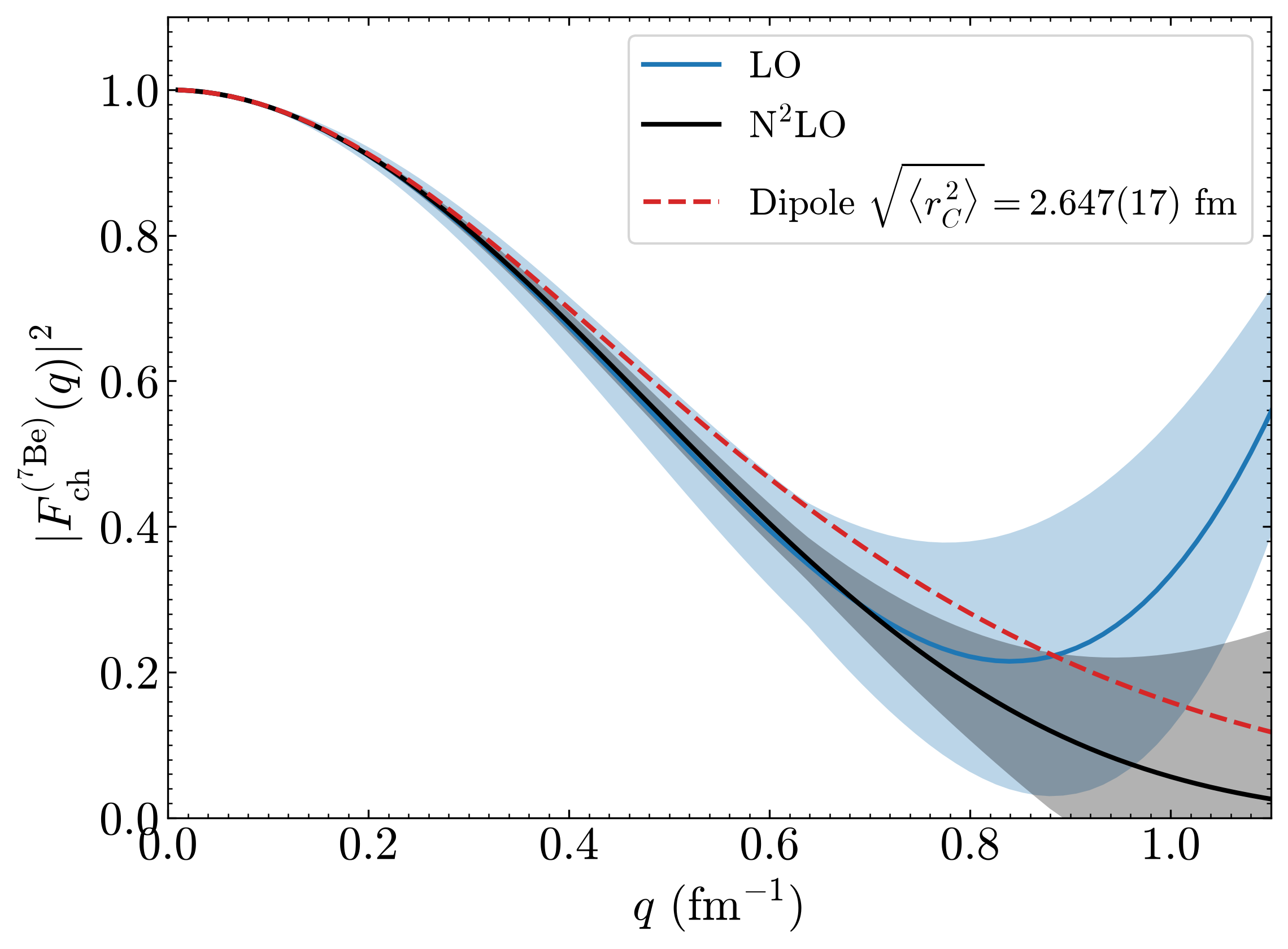}
    \includegraphics[width=.32\textwidth]{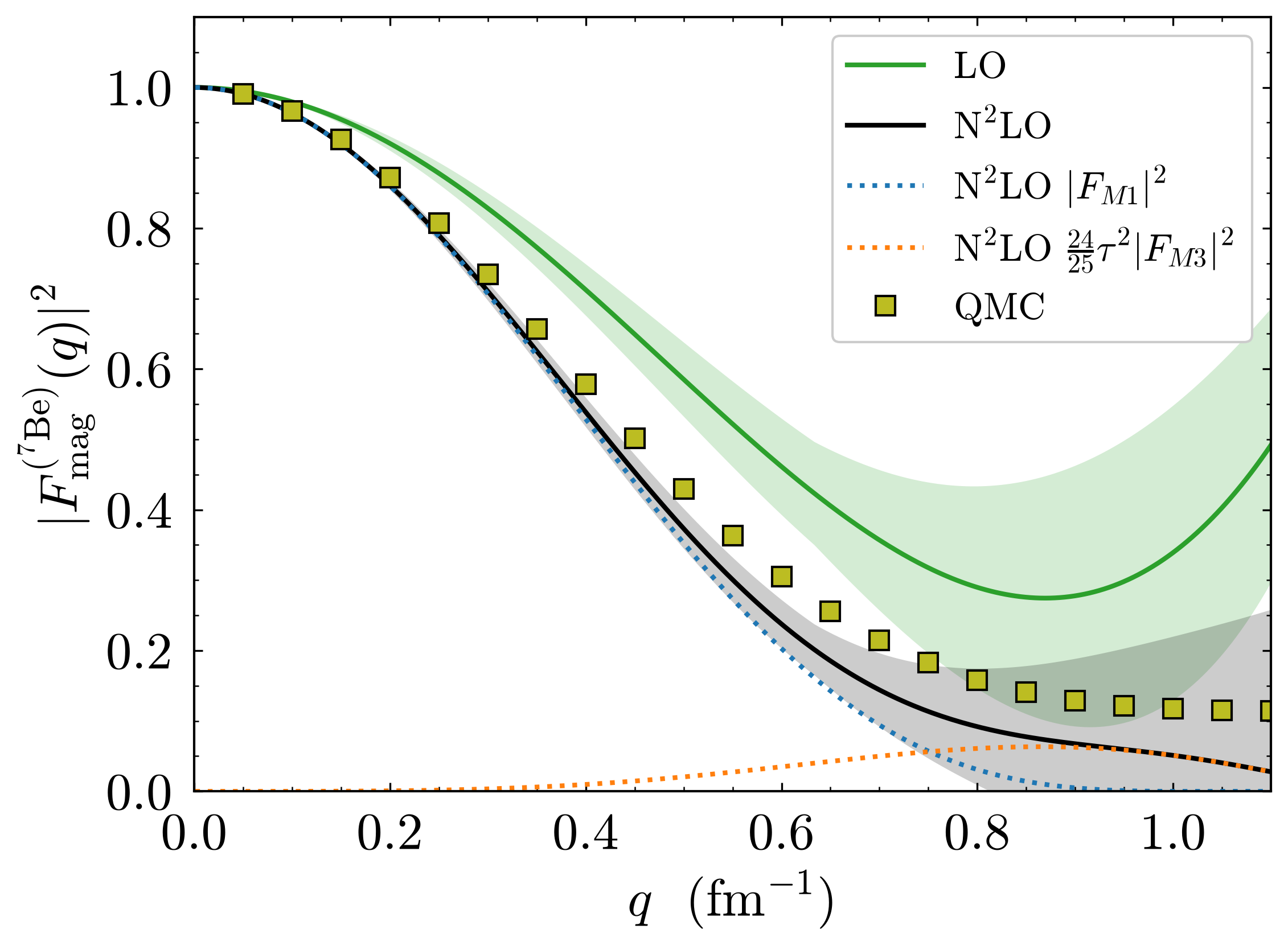}
    \includegraphics[width=.32\textwidth]{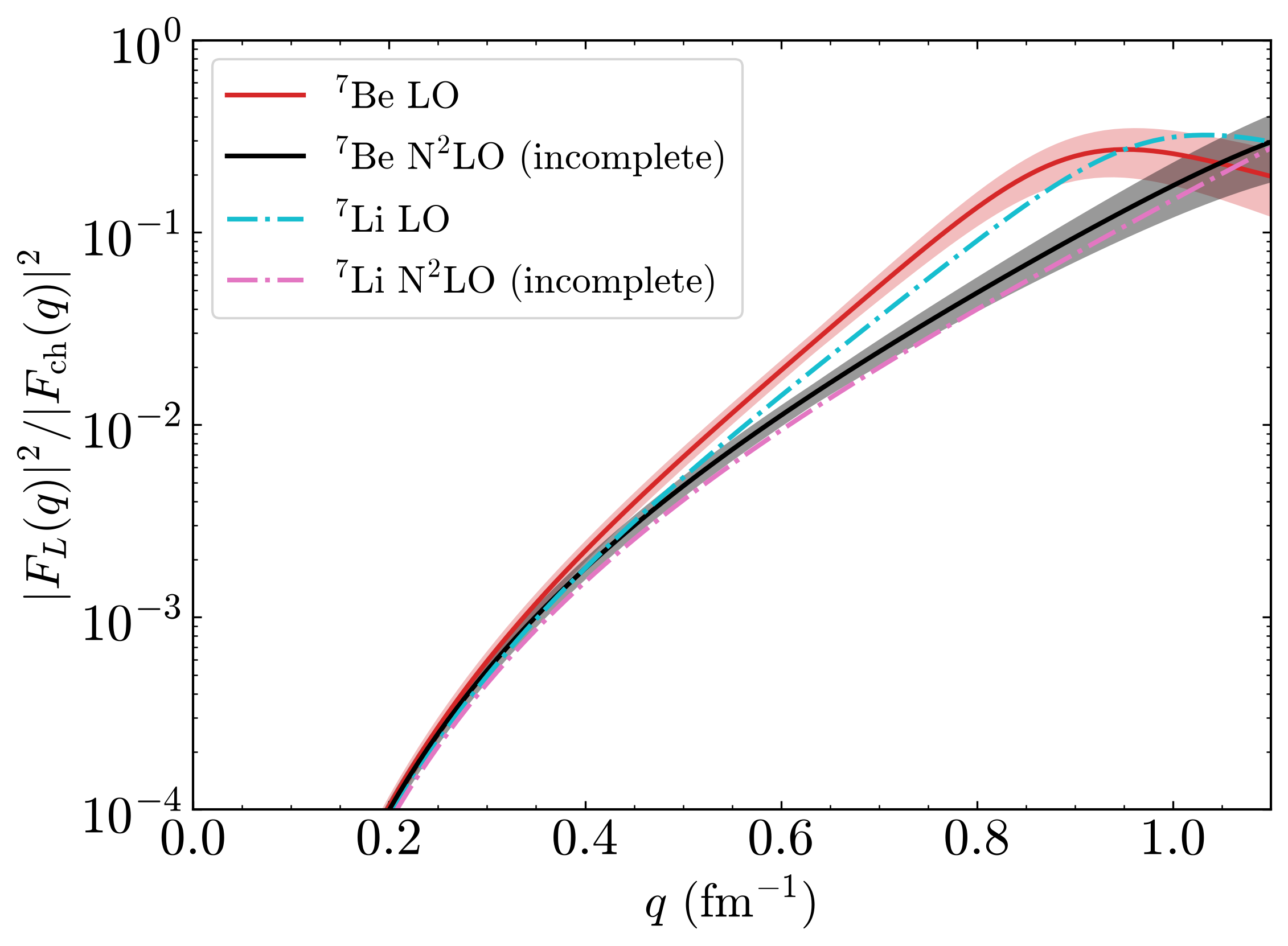}
    \caption{Plot of \Be's squared charge form factor (left), magnetic form factor (middle) and the ratio of the longitudinal form factor (right) as a function of momentum transfer $q$.  }
    \label{fig:Beseven_charge_mag}
\end{figure}
We then calculate the charge form factor and the ratio of the longitudinal form factor for the first excited state to the charge form factor (see \cref{fig:Beseven_charge_mag}).   Instead, we can calculate the reduced electric transition probability $B(\rm E2,3/2\to1/2)$ at LO using \cref{eq:BEtwoQ},
\begin{equation}
    B({\rm E2})=\frac{25}{16\pi}\left[\quadbe\right]^2=17.0\pm(0.9)_{\rm exp}\pm(2.2)_{\rm theo}~\text{fm}^4. 
\end{equation}
This value is consistent with a cluster model study in Ref.~\cite{Mason:2008ka}, which calculated $B({\rm E2})=18.3$ \unit{fm^4}. A more detailed comparison is provided in Tab.~\ref{tab:Beseven} below. 
\setlength{\tabcolsep}{5pt}
\renewcommand{\arraystretch}{1.3}
\begin{table}[h]
    \centering
    \begin{tabular}{l|c|c}\hline\hline
       Method & $B({\rm E2})$ & Ref. \\\hline
       Cluster model & 18.3 & \cite{Mason:2008ka}\\
        Nuclear lattice EFT & $15.2(5), ~16.0(2)$ &\cite{Shen:2024qzi}\\
        NCSMC & $20.02$ &\cite{Vorabbi:2019imi}\\
        Ab initio GFMC & $25.6(3),~27.5(8)$  &\cite{Pervin:2007sc}\\
        & 22.2(11) &\cite{PhysRevC.87.035503} \\
        Microscopic cluster model & 26.2 & \cite{Tao:2025ukl}\\
       Cluster EFT & $17.0(2.4)_{\rm comb}$ &  This work\\\hline
        Experiment & $26(6)(3)$ &\cite{Henderson:2019ubp}\\\hline\hline
    \end{tabular}
    \caption{Reduced transition probability $B({\rm E2})$ of \Be. All values are given in \unit{e^2fm^4}.}
    \label{tab:Beseven}
\end{table}

Finally, we can take advantage of \cref{eq:BEtwoQrelation} to estimate
\begin{equation}
\frac{\Cpd'}{\Cpq'}=\frac{\xi_2}{\xi_2'}=0.904,
\end{equation}
s in close agreement with the NCSMC value $0.903$~\cite{Atkinson:2024zrm}. It follows that
\begin{equation}
\Cpd'=\Cpq'\sqrt{\left(\frac{\xi_2}{\xi_2'}\right)^2}=3.41\pm(0.05)_{\rm exp}\pm(0.04)_{\rm theo}~\text{fm}^{-1/2}.
\end{equation}

\section{Conclusions \label{sec:conlusion}}

Few-nucleon systems are perfect for testing our understanding of nuclear forces and gaining new insights into non-perturbative quantum chromodynamics in low-energy regimes. Coulomb effects at low energies can be calculated non-perturbatively. Cluster EFT provides a model-independent and systematic framework to study nuclei with a cluster structure by taking advantage of the separation of scales and treating these clusters as fundamental degrees of freedom. Our work in this paper focuses on the EM properties of two systems, $\alpha$-triton and $\alpha$-helion, which are directly connected to the nuclear structures of \Liseven and \Be, respectively. Cluster EFT enables robust correlations between key quantities, such as the asymptotic normalization coefficients, charge radii, magnetic dipole moments and other observables within a unified framework, which are helpful to address the longstanding discrepancies in theoretical and empirical determinations of ANCs for $\alpha + t\to \Liseven$ and $\alpha + h \to\Be$ reactions. Our derivation reveals richer connections among these quantities that may have been disregarded in previous studies. Using these relationships and the measured charge radii of \Liseven, \Be, and their constituent clusters, we determine the ANCs with few-percent uncertainties. We have demonstrated that cluster EFT successfully captures the essential physics of \Liseven at low energies: the quadrupole moment and $B({\rm E2})$ serve as benchmarks for the breakdown scale, and the convection current of the constituents is essential to reproduce the magnetic moments. This is encouraging for its application in describing the nuclear structure of \Be, such as its quadrupole moment, reduced electric quadrupole transition probability, magnetic radius, and the ratio of the ANCs of its ground and first excited states. These results can serve as valuable references for future experiments and theoretical calculations. The EM form factors of these nuclei have also been calculated. For \Liseven they are in good agreement with the experimental data for momentum transfers up to $1$~\unit{fm^{-1}}, and the magnetic form factor of \Be agrees with quantum Monte Carlo calculations.

We note that the approach is less suited to Lithium-6 nucleus, where the deuteron breakup lies only $2.2$~MeV above the $\alpha+d$ threshold, so that no comparable separation of scales exists. A description in terms of $\alpha+n+p$ degrees of freedom would be required there. Lastly, this work underscores the effectiveness of the rigorous yet straightforward approach provided by the cluster EFT framework in illuminating the significant degrees of freedom in complex systems. Our findings can serve as a benchmark for future extensions of cluster EFT to multi-body systems, enhancing our understanding of nuclear dynamics across a wide range of energy scales.

\section{Acknowledgments}
We thank R.~P.~Springer for her mentorship, encouragement, and stimulating discussions throughout the course of this work. We thank R.~P.~Springer and T.~R.~Richardson for comments on the manuscript.  We are grateful to G.~King and collaborators for making their data accessible. We thank E.~M.~Tursunov for pointing out an error in an earlier version of this manuscript and for their valuable feedback. We acknowledge the use of AI assistance, specifically ChatGPT \cite{ChatGPT} during the preparation of an earlier version of this manuscript and Claude (Anthropic) \cite{claude-opus-5} in refining the language and clarity of this manuscript. The initial stages of this work were completed during the author's time at Duke University. This work was supported in part by the U.S. Department of Energy, Office of Science, Nuclear Physics program under Award No. DE-FG02-05ER41368 and the Lenfest Grant program at Washington and Lee University.

\appendix

\section{Electromagnetic current matrix elements in non-relativistic limit \label{app:em_current}}
\subsection{$\frac{3}{2}\to \frac{3}{2}$ electromagnetic matrix element}
The electromagnetic interaction current for spin-3/2 particles has the following general simple form
\begin{equation}
    J^\mu = -\Bar{u}_\alpha(p',x')\left\{g^{\alpha\beta}\left[\gamma^\mu F_1(Q^2) +\frac{i\sigma^{\mu\nu}q_\nu}{2M}F_2(Q^2)\right]-\frac{q^\alpha q^\beta}{2M^2}\left[\gamma^\mu F_3(Q^2) +\frac{i\sigma^{\mu\nu}q_\nu}{2\msev}F_4(Q^2)\right]\right\}u_\beta(p,x),
\end{equation} 
where $u_\alpha(p,M)$ is the Rarita-Schwinger spinor expressed in terms of the spin-1 vector $e_\alpha(p,m )$ and the Dirac spinor $u(p,s)$
\cite{Ericson:1988gk},
\begin{align}
    u_\mu(p,x) &=\sum_{s,m}\left\langle 1m\frac{1}{2}s\big|\frac{3}{2}x\right\rangle \epsilon_\mu(p,m)u(p,s)\equiv \left[S_\mu\right]_{xs}u(p,s),\\
    u(p,s) & =\sqrt{M+E}\begin{pmatrix}
        1\\
        \frac{\vb*{\sigma}\cdot\vb{p} }{M+E}
    \end{pmatrix}\chi(s),\\
        \epsilon_\mu(p,m) & =\left(\frac{\vb*{\epsilon}_m\cdot \vb{p}}{M},\vb*{\epsilon}_m + \frac{\vb{p}~[\vb*{\epsilon}_m\cdot \vb{p}]}{M(M+E)}\right)
\end{align}
In the first expression, $S_\alpha$'s are $2\times 4$ spin transition matrices connecting states with total angular momentum $j = 1/2$ and $ j = 3/2$. 
In the non-relativistic limit, this reduces to
\begin{equation}
  \va{\chi}\left(\frac{3}{2},x\right) =   \sum_{s,m}\left\langle 1m\frac{1}{2}s\big|\frac{3}{2}x\right\rangle \vb*{\epsilon}_m\chi(s)\equiv\left[\vb{S}\right]_{xs}\chi(s),
\end{equation}
where 
\begin{equation}
    \vb{S}^\dagger = \sum_m \left\langle 1m\frac{1}{2}s\big|\frac{3}{2}x\right\rangle \vb*{\epsilon}_m.
\end{equation}
It follows that
\begin{equation}
    {S}_\mu = \left(\frac{\vb{S} \cdot \vb{p}}{M},\vb{S} + \frac{\vb{p}~[\vb{S}\cdot \vb{p}]}{M(M+E)}\right).
\end{equation}
The states $\vec{\chi}\left(3/2,x\right)$ are by definition three-vectors, each component of which is a spinor. The transition matrix for particle spin-3/2 to spin-1/2 is given as following,
\begin{equation}
S_\mu=\begin{pmatrix}
\epsilon_\mu(p,+1) & \sqrt{\frac{2}{3}}\epsilon_\mu(p,0)&\sqrt{\frac{1}{3}} \epsilon_\mu(p,-1) & 0\\
0 & \sqrt{\frac{1}{3}}\epsilon_\mu(p,+1) & \sqrt{\frac{2}{3}}\epsilon_\mu(p,0) & \epsilon_\mu(p,-1)
\end{pmatrix}
\end{equation}
We obtain
\begin{align}
    S_1 &= \frac{1}{\sqrt{6}}\begin{pmatrix}
-{\sqrt{3}} & 0 & 1 & 0\\
0 & -1 & 0 &\sqrt{3}
\end{pmatrix},\\
  S_2 &= -\frac{i}{\sqrt{6}}\begin{pmatrix}
{\sqrt{3}} & 0 & 1 & 0\\
0 & 1 & 0 &\sqrt{3}
\end{pmatrix},\\
  S_3 &= \frac{2}{\sqrt{6}}\begin{pmatrix}
0 & 1 & 0 & 0\\
0 & 0 & 1 & 0
\end{pmatrix}.
\end{align}
They satisfy
\begin{align}
S_i S_i^\dagger &= \frac{2}{3}\delta_{ij} - \frac{i}{3}\epsilon_{ijk}\sigma_k,\\
S_i^\dagger S_j &= \frac{3}{4}\delta_{ij} - \frac{1}{6}\left\{J^{\left(3/2\right)}_i,J^{\left(3/2\right)}_j\right\} + \frac{i}{3}\epsilon_{ijk}J^{\left(3/2\right)},
\end{align}
where $J^{(3/2)}_i$’s are the generators of the spin-3/2. Using the identity $S_i =i\epsilon_{ijk}\sigma_j S_k$ it can be shown that
\begin{align}
    S_i (\vb*{\sigma}\times \vb{q})^k S_i =\frac{2}{3} \left(\vb{J}^{(3/2)}\times \vb{q}\right)^k.
\end{align}
The matrix elements of the electromagnetic currents in the non-relativistic limit become
\begin{equation}
    \langle \vb{p}', x|J^0_\text{em}|\vb{p},y\rangle = e\left\{(F_1-\tau F_2)\delta_{xy} + \frac{1}{2(\msev)^2} \left(\vb{S}^\dagger\cdot \vb{q}\right)\left(\vb{S}\cdot \vb{q}\right)\left[(F_1-\tau F_2) + (1+\tau)(F_3-\tau F_4)\right]  \right\},
\end{equation}
where $\tau=\vb{q}^2/4(\msev)^2$. The electric multipole form factors are defined as
\begin{align}
    F_{\rm C0}(q)-\frac{2}{3}\tau F_{\rm C2} &= F_1-\tau F_2,\\
    F_{\rm C2}(q) &= F_1-\tau F_2+(1+\tau)[F_3-\tau F_4].
\end{align}
Similarly, the vector component of the electromagnetic current reduces to
\begin{equation}
\begin{aligned}
    \langle \vb{p}', x|J^k_\text{em}|\vb{p},y\rangle = \frac{e}{2\msev}&\bigg\{(F_1+F_2)S_i^\dagger\left[i\left(\vb*{\sigma}\times \vb{q}\right)^k\right]S_i \\
    &+ \frac{1}{2(\msev)^2} \left(\vb{S}^\dagger\cdot \vb{q}\right)\left[(F_1+F_2) + (1+\tau)(F_3+ F_4)\right]\left[i\left(\vb*{\sigma}\times \vb{q}\right)^k\right]\left(\vb{S}\cdot \vb{q}\right)  \bigg\}.
    \end{aligned}
\end{equation} 
The magnetic multipole form factors are defined as
\begin{align}
    F_{\rm M1}(q)-\frac{4}{5}\tau F_{\rm M3}(q) &= F_1+ F_2,\\
    F_{\rm M3}(q) &= F_1+ F_2+(1+\tau)[F_3+ F_4].
\end{align}

\section{Partial-wave expansion of Coulomb Green’s function \label{app:Green_partial_wave}}
\begin{figure}[h]
    \centering
    \includegraphics[width=0.2\linewidth]{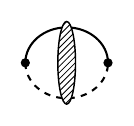}
    \caption{The irreducible self-energy diagram with Coulomb photon exchange.}
    \label{fig:coulomb_green}
\end{figure}

For low momenta the Coulomb interaction must be included to all orders and we do this by using the full Coulomb propagator $G_C$, shown in Fig.~\ref{fig:coulomb_green}. The Coulomb propagator, or Coulomb Green’s function, can be written in the spectral representation as

\begin{equation}
(\vb{r}'|G_C|\vb{r})=\int\frac{\dd^3\vb{p}}{(2\pi)^3}\frac{\psi^{(+)}_{\vb{p}}(\vb{r}')\psi^{(+)*}_{\vb{p}}(\vb{r})}{E-\frac{\vb{p}^2}{2m_R}+i\epsilon},
\end{equation}
where we have represented it in coordinate space. We define the Coulomb wavefunction $\psi_{\vb{p}}(\vb{r})$ by its partial wave expansion \cite{Landau:1991wop},
\begin{equation}
    \psi_{\vb{p}}^{(\pm)}(\vb{r})=\sum_{\ell,m}4\pi i^\ell e^{\pm i\sigma_\ell}\frac{F_\ell(\eta,pr)}{pr}Y^m_\ell(\vu{p})Y^{m*}_\ell(\vu{r}),
\end{equation}
with $F_\ell(\eta,pr)$ being the regular Coulomb wave function given by
\begin{equation}
    F_\ell(\eta,pr)=\frac{C_{\eta,\ell}}{2^{\ell+1}}(\mp i)^{\ell+1}M_{\pm i\eta,\ell+1/2}(\pm 2ipr).
\end{equation}
Here $M_{a,b}(z)$ is the Whittaker M-function \cite[\href{https://dlmf.nist.gov/13.14.2}{(13.14.2)}]{NIST:DLMF}. It is useful to express the Green’s function in its partial-wave expanded form
\begin{equation}
    (\vb{r}'|G_C|\vb{r}) = \sum_{\ell, m}4\pi G^{(\ell)}_C(E,r',r) Y^m_\ell(\hatr')Y^{m*}_\ell(\hatr).
    \label{eq:coulomb_green_pw1}
\end{equation}
Expanding the left hand site of the above equation using the definition of the Coulomb Green function, we see that
\begin{equation}
\begin{aligned}
 (\vb{r}'|G_C|\vb{r}) &=  \sum_{\ell,m,\ell',m'}(4\pi)^2 i^{(\ell'-\ell)}e^{i(\sigma_{\ell'}-\sigma_\ell)}\int\frac{\dd^3\vb{p}}{(2\pi)^3}\frac{\frac{F_{\ell'}(\eta, pr')}{pr'}\frac{F^*_{\ell}(\eta, pr)}{pr}}{E-\frac{\vb{p}^2}{2m_R}+i\epsilon}\\
 &\qquad\times Y^{m'}_{\ell'}(\hat{\vb{p}})Y^{m'*}_{\ell'}(\hatr') Y^{m*}_\ell(\hat{\vb{p}})Y^{m}_\ell(\hatr)\\
 &=\sum_{\ell,m}4\pi \int\frac{\dd^3\vb{p}}{(2\pi)^3}\frac{\frac{F_{\ell}(\eta, pr')}{pr'}\frac{F^*_{\ell}(\eta, pr)}{pr}}{E-\frac{\vb{p}^2}{2m_R}+i\epsilon}Y^{m}_\ell(\hatr')Y^{m*}_\ell(\hatr)
\end{aligned}  
\label{eq:coulomb_green_pw2}
\end{equation}
Comparing Eqs.~\eqref{eq:coulomb_green_pw1} and ~\eqref{eq:coulomb_green_pw2}, we deduce that the Green’s function for a specific partial wave is given by
\begin{equation}
 G^{(\ell)}_C(E,r',r) =  \int\frac{\dd^3\vb{p}}{(2\pi)^3}\frac{F_{\ell}(\eta, pr')}{pr'}\frac{F^*_{\ell}(\eta, pr)}{pr}\frac{1}{E-\frac{\vb{p}^2}{2m_R}+i\epsilon}
\end{equation}
It is sometimes convenient to use the Coulomb Green’s function in a non-integral form andwe present such a form for the bound-state Green’s function below (see Ref.~\cite{Ryberg:2015lea} for more details) 
\begin{equation}
  G^{(\ell)}_C(-B,\rho',\rho) =   -\frac{m_Rp}{2\pi} \frac{F_\ell(\eta,\rho')}{\rho'}\frac{e^{i\sigma_\ell}e^{(\eta-i\ell)\pi/2}W_{-\frac{k_C}{\gamma},\ell+\frac{1}{2}}(2\gamma r)}{\rho},
\end{equation}
where $\rho = i\gamma r$ with $\gamma = \sqrt{2m_RB}$ being the binding momentum. Many of the diagrams we consider will have a factor of a Coulomb Green’s function with one end at zero separation. Using the idendity 
\begin{equation}
    \lim_{r\rightarrow0}\frac{F_\ell(\eta,pr)}{(pr)^{\ell+1}} =C_{\eta,\ell} = \frac{2^\ell e^{-\frac{\pi\eta}{2}}|\Gamma(\ell+1+i\eta)|}{\Gamma(2\ell+2)}
\end{equation}
It follows that 
\begin{equation}
    \lim_{r'\rightarrow0}\frac{G^{(\ell)}(-B,r',r)}{r'^\ell}=-\frac{m_R(i\gamma)^\ell}{2\pi}\frac{2^\ell e^{i\left(\sigma_\ell-\frac{\ell\pi}{2}\right)}|\Gamma(\ell+1+i\eta)|}{\Gamma(2\ell+2)}\frac{W_{-\frac{k_C}{\gamma},\ell+\frac{1}{2}}(2\gamma r)}{r}
    \label{eq:Green_Swave}
\end{equation}
This result will be used to solve the loop-integrals for the quadrupole moment numerically.


\bibliography{ref.bib}

@article{Chambers-Wall:2024fha,
    author = "Chambers-Wall, G. and Gnech, A. and King, G. B. and Pastore, S. and Piarulli, M. and Schiavilla, R. and Wiringa, R. B.",
    title = "{Quantum Monte~Carlo Calculations of Magnetic Form Factors in Light Nuclei}",
    eprint = "2407.03487",
    archivePrefix = "arXiv",
    primaryClass = "nucl-th",
    doi = "10.1103/PhysRevLett.133.212501",
    journal = "Phys. Rev. Lett.",
    volume = "133",
    number = "21",
    pages = "212501",
    year = "2024"
}

@book{Landau:1991wop,
    author = "Landau, Lev Davidovich and Lifshits, E. M.",
    title = "{Quantum Mechanics}: {Non-Relativistic Theory}",
    doi = "10.1016/C2013-0-02793-4",
    isbn = "978-0-7506-3539-4",
    publisher = "Butterworth-Heinemann",
    address = "Oxford",
    series = "Course of Theoretical Physics",
    volume = "v.3",
    year = "1991"
}

@article{NevoDinur:2018hdo,
    author = "Nevo Dinur, N. and Hernandez, O. J. and Bacca, S. and Barnea, N. and Ji, C. and Pastore, S. and Piarulli, M. and Wiringa, R. B.",
    title = "{Zemach moments and radii of $^{2,3}$H and $^{3,4}$He}",
    eprint = "1812.10261",
    archivePrefix = "arXiv",
    primaryClass = "nucl-th",
    reportNumber = "INT-PUB-18-064, MITP/18-128",
    doi = "10.1103/PhysRevC.99.034004",
    journal = "Phys. Rev. C",
    volume = "99",
    number = "3",
    pages = "034004",
    year = "2019"
}

@article{Chen:1999tn,
	title        = {{Nucleon-nucleon effective field theory without pions}},
	author       = {Chen, Jiunn-Wei and Rupak, Gautam and Savage, Martin J.},
	year         = 1999,
	journal      = {Nucl. Phys.},
	volume       = {A653},
	pages        = {386--412},
	doi          = {10.1016/S0375-9474(99)00298-5},
	eprint       = {nucl-th/9902056},
	archiveprefix = {arXiv},
	primaryclass = {nucl-th},
	reportnumber = {NT-UW-99-14, JLAB-THY-99-50}
}

@article{Kaplan_KSW_1998b,
	title        = {{A New expansion for nucleon-nucleon interactions}},
	author       = {Kaplan, David B. and Savage, Martin J. and Wise, Mark B.},
	year         = 1998,
	journal      = {Phys. Lett.},
	volume       = {B424},
	pages        = {390--396},
	doi          = {10.1016/S0370-2693(98)00210-X},
	eprint       = {nucl-th/9801034},
	archiveprefix = {arXiv},
	primaryclass = {nucl-th},
	reportnumber = {DOE-ER-40561-352, INT-97-00-189, NT-UW-98-05, CALT-68-2155}
}

@article{Kaplan_KSW_1998a,
	title        = {{Two nucleon systems from effective field theory}},
	author       = {Kaplan, David B. and Savage, Martin J. and Wise, Mark B.},
	year         = 1998,
	journal      = {Nucl. Phys.},
	volume       = {B534},
	pages        = {329--355},
	doi          = {10.1016/S0550-3213(98)00440-4},
	eprint       = {nucl-th/9802075},
	archiveprefix = {arXiv},
	primaryclass = {nucl-th},
	reportnumber = {DOE-ER-40561-357, INT-98-00-5, NT-UW-98-08, CALT-68-2161}
}

@article{Kaplan_KSW_1996,
	title        = {{Nucleon - nucleon scattering from effective field theory}},
	author       = {Kaplan, David B. and Savage, Martin J. and Wise, Mark B.},
	year         = 1996,
	journal      = {Nucl. Phys. B},
	volume       = 478,
	pages        = {629--659},
	doi          = {10.1016/0550-3213(96)00357-4},
	eprint       = {nucl-th/9605002},
	archiveprefix = {arXiv},
	reportnumber = {DOE-ER-40561-257, INT-96-00-125, UW-PT-96-06, CMU-HEP-96-06, DOE-ER-40862-117, CALT-68-2047}
}

@article{Hammer:2019poc,
	title        = {{Nuclear effective field theory: status and perspectives}},
	author       = {Hammer, H.-W. and K\"onig, S. and van Kolck, U.},
	year         = 2020,
	journal      = {Rev. Mod. Phys.},
	volume       = 92,
	number       = 2,
	pages        = {025004},
	doi          = {10.1103/RevModPhys.92.025004},
	eprint       = {1906.12122},
	archiveprefix = {arXiv},
	primaryclass = {nucl-th}
}

@article{Kong:1999sf,
	title        = {{Coulomb effects in low-energy proton proton scattering}},
	author       = {Kong, Xinwei and Ravndal, Finn},
	year         = 2000,
	journal      = {Nucl. Phys. A},
	volume       = 665,
	pages        = {137--163},
	doi          = {10.1016/S0375-9474(99)00406-6},
	eprint       = {hep-ph/9903523},
	archiveprefix = {arXiv}
}

@article{Chen:1999vd,
    author = "Chen, Jiunn-Wei and Rupak, Gautam and Savage, Martin J.",
    title = "{Suppressed amplitudes in n p ---{\ensuremath{>}} d gamma}",
    eprint = "nucl-th/9905002",
    archivePrefix = "arXiv",
    reportNumber = "NT-UW-99-20",
    doi = "10.1016/S0370-2693(99)01007-2",
    journal = "Phys. Lett. B",
    volume = "464",
    pages = "1--11",
    year = "1999"
}

@article{Tao:2025ukl,
    author = "Tao, De-Ye and Zhou, Bo and Wang, Si-Min and Ma, Yu-Gang",
    title = "{Analysis of clustering fragments of Li7 and Be7 in the microscopic cluster model}",
    eprint = "2501.05216",
    archivePrefix = "arXiv",
    primaryClass = "nucl-th",
    doi = "10.1103/PhysRevC.111.014323",
    journal = "Phys. Rev. C",
    volume = "111",
    number = "1",
    pages = "014323",
    year = "2025"
}

@article{Chambers-Wall:2024uhq,
    author = "Chambers-Wall, G. and Gnech, A. and King, G. B. and Pastore, S. and Piarulli, M. and Schiavilla, R. and Wiringa, R. B.",
    title = "{Magnetic structure of A{\ensuremath{\leq}}10 nuclei using the Norfolk nuclear models with quantum Monte Carlo methods}",
    eprint = "2407.04744",
    archivePrefix = "arXiv",
    primaryClass = "nucl-th",
    doi = "10.1103/PhysRevC.110.054316",
    journal = "Phys. Rev. C",
    volume = "110",
    number = "5",
    pages = "054316",
    year = "2024"
}

@article{Stellin:2020gst,
	title        = {{P-Wave Two-Particle Bound and Scattering States in a Finite Volume including QED}},
	author       = {Stellin, Gianluca and Mei\ss{}ner, Ulf-G.},
	year         = 2021,
	journal      = {Eur. Phys. J. A},
	volume       = 57,
	number       = 1,
	pages        = 26,
	doi          = {10.1140/epja/s10050-020-00319-1},
	eprint       = {2008.06553},
	archiveprefix = {arXiv},
	primaryclass = {hep-lat}
}

@article{Ryberg:2015lea,
	title        = {{Range corrections in Proton Halo Nuclei}},
	author       = {Ryberg, Emil and Forss\'en, Christian and Hammer, H. -W. and Platter, Lucas},
	year         = 2016,
	journal      = {Annals Phys.},
	volume       = 367,
	pages        = {13--32},
	doi          = {10.1016/j.aop.2016.01.008},
	eprint       = {1507.08675},
	archiveprefix = {arXiv},
	primaryclass = {nucl-th}
}

@article{Igamov:2007svc,
    author = "Igamov, S. B. and Yarmukhamedov, R.",
    title = "{Modified two-body potential approach to the peripheral direct capture astrophysical 					a					+					A					\textrightarrow{}					B					+					\ensuremath{\gamma}				 reaction and asymptotic normalization coefficients}",
    doi = "10.1016/j.nuclphysa.2006.10.041",
    journal = "Nucl. Phys. A",
    volume = "781",
    pages = "247--276",
    year = "2007",
    note = "[Erratum: Nucl.Phys.A 832, 346--347 (2010)]"
}

@article{IGAMOV2010346,
author = {S.B. Igamov and R. Yarmukhamedov},
title = {Erratum to “Modified two-body potential approach to the peripheral direct capture astrophysical a+A→B+γ reaction and asymptotic normalization coefficients” [Nucl. Phys. A 781 (2007) 247]},
journal = {Nuclear Physics A},
volume = {832},
number = {3},
pages = {346-347},
year = {2010},
issn = {0375-9474},
doi = {https://doi.org/10.1016/j.nuclphysa.2009.12.004},
url = {https://www.sciencedirect.com/science/article/pii/S0375947409009336},
}

@article{RevModPhys.83.195,
  title = {Solar fusion cross sections. II. The $pp$ chain and CNO cycles},
  author = {Adelberger, E. G. and Garc\'{\i}a, A. and Robertson, R. G. Hamish and Snover, K. A. and Balantekin, A. B. and Heeger, K. and Ramsey-Musolf, M. J. and Bemmerer, D. and Junghans, A. and Bertulani, C. A. and Chen, J.-W. and Costantini, H. and Prati, P. and Couder, M. and Uberseder, E. and Wiescher, M. and Cyburt, R. and Davids, B. and Freedman, S. J. and Gai, M. and Gazit, D. and Gialanella, L. and Imbriani, G. and Greife, U. and Hass, M. and Haxton, W. C. and Itahashi, T. and Kubodera, K. and Langanke, K. and Leitner, D. and Leitner, M. and Vetter, P. and Winslow, L. and Marcucci, L. E. and Motobayashi, T. and Mukhamedzhanov, A. and Tribble, R. E. and Nollett, Kenneth M. and Nunes, F. M. and Park, T.-S. and Parker, P. D. and Schiavilla, R. and Simpson, E. C. and Spitaleri, C. and Strieder, F. and Trautvetter, H.-P. and Suemmerer, K. and Typel, S.},
  journal = {Rev. Mod. Phys.},
  volume = {83},
  issue = {1},
  pages = {195--245},
  numpages = {0},
  year = {2011},
  month = {Apr},
  publisher = {American Physical Society},
  doi = {10.1103/RevModPhys.83.195},
  url = {https://link.aps.org/doi/10.1103/RevModPhys.83.195}
}

@article{Mohr:2009uz,
    author = "Mohr, Peter",
    title = "{Low-energy He-3(alpha,alpha) He-3 elastic scattering and the He-3(alpha,gamma) Be-7 reaction}",
    eprint = "0906.3000",
    archivePrefix = "arXiv",
    primaryClass = "nucl-th",
    doi = "10.1103/PhysRevC.79.065804",
    journal = "Phys. Rev. C",
    volume = "79",
    pages = "065804",
    year = "2009"
}

@article{PhysRevC.89.014325,
  title = {Effective field theory for proton halo nuclei},
  author = {Ryberg, Emil and Forss\'en, Christian and Hammer, H.-W. and Platter, Lucas},
  journal = {Phys. Rev. C},
  volume = {89},
  issue = {1},
  pages = {014325},
  numpages = {6},
  year = {2014},
  month = {Jan},
  publisher = {American Physical Society},
  doi = {10.1103/PhysRevC.89.014325},
  url = {https://link.aps.org/doi/10.1103/PhysRevC.89.014325}
}

@article{Krauth:2021foz,
	title        = {{Measuring the \ensuremath{\alpha}-particle charge radius with muonic helium-4 ions}},
	author       = {Krauth, Julian J. and others},
	year         = 2021,
	journal      = {Nature},
	volume       = 589,
	number       = 7843,
	pages        = {527--531},
	doi          = {10.1038/s41586-021-03183-1}
}

@article{Hammer:2017tjm,
	title        = {{Effective field theory description of halo nuclei}},
	author       = {Hammer, H. -W. and Ji, C. and Phillips, D. R.},
	year         = 2017,
	journal      = {J. Phys. G},
	volume       = 44,
	number       = 10,
	pages        = 103002,
	doi          = {10.1088/1361-6471/aa83db},
	eprint       = {1702.08605},
	archiveprefix = {arXiv},
	primaryclass = {nucl-th}
}

@article{Sick:2014yha,
    author = "Sick, Ingo",
    title = "{Zemach moments of $^3$He and $^4$He}",
    eprint = "1412.2603",
    archivePrefix = "arXiv",
    primaryClass = "nucl-ex",
    doi = "10.1103/PhysRevC.90.064002",
    journal = "Phys. Rev. C",
    volume = "90",
    number = "6",
    pages = "064002",
    year = "2014"
}

@article{Mason:2008ka,
    author = "Mason, A. and Chatterjee, R. and Fortunato, L. and Vitturi, A.",
    title = "{Electric and magnetic response to the continuum for A=7 isotopes in a dicluster model}",
    eprint = "0806.4032",
    archivePrefix = "arXiv",
    primaryClass = "nucl-th",
    doi = "10.1140/epja/i2008-10685-3",
    journal = "Eur. Phys. J. A",
    volume = "39",
    pages = "107--116",
    year = "2009"
}

@article{Henderson:2019ubp,
    author = "Henderson, S. L. and others",
    title = "{First measurement of the B(E2;3/2\ensuremath{-}\textrightarrow{}1/2\ensuremath{-} ) transition strength in Be7 : Testing abinitio predictions for A=7 nuclei}",
    eprint = "2109.07312",
    archivePrefix = "arXiv",
    primaryClass = "nucl-ex",
    doi = "10.1103/PhysRevC.99.064320",
    journal = "Phys. Rev. C",
    volume = "99",
    number = "6",
    pages = "064320",
    year = "2019"
}

@article{Buck:1977xyz,
    author = "Buck, B. and Pilt, A. A.",
    title = "{Alpha-particle and triton cluster states in 19 F}",
    doi = "10.1016/0375-9474(77)90300-1",
    journal = "Nucl. Phys. A",
    volume = "280",
    pages = "133--160",
    year = "1977"
}

@article{RevModPhys.93.025010,
  title = {CODATA recommended values of the fundamental physical constants: 2018},
  author = {Tiesinga, Eite and Mohr, Peter J. and Newell, David B. and Taylor, Barry N.},
  journal = {Rev. Mod. Phys.},
  volume = {93},
  issue = {2},
  pages = {025010},
  numpages = {63},
  year = {2021},
  month = {Jun},
  publisher = {American Physical Society},
  doi = {10.1103/RevModPhys.93.025010},
  url = {https://link.aps.org/doi/10.1103/RevModPhys.93.025010}
}

@article{PhysRevC.100.054307,
	title        = {Low-energy bound states, resonances, and scattering of light ions},
	author       = {Luna, Benjamin K. and Papenbrock, T.},
	year         = 2019,
	month        = {Nov},
	journal      = {Phys. Rev. C},
	publisher    = {American Physical Society},
	volume       = 100,
	pages        = {054307},
	doi          = {10.1103/PhysRevC.100.054307},
	url          = {https://link.aps.org/doi/10.1103/PhysRevC.100.054307},
	issue        = 5,
	numpages     = 17
}

@article{Shen:2024qzi,
    author = "Shen, Shihang and Elhatisari, Serdar and Lee, Dean and Mei{\ss}ner, Ulf-G. and Ren, Zhengxue",
    title = "{Ab~Initio Study of the Beryllium Isotopes Be7 to Be12}",
    eprint = "2411.14935",
    archivePrefix = "arXiv",
    primaryClass = "nucl-th",
    doi = "10.1103/PhysRevLett.134.162503",
    journal = "Phys. Rev. Lett.",
    volume = "134",
    number = "16",
    pages = "162503",
    year = "2025"
}

@article{Vorabbi:2019imi,
    author = "Vorabbi, Matteo and Navr\'atil, Petr and Quaglioni, Sofia and Hupin, Guillaume",
    title = "{$^7$Be and $^7$Li nuclei within the no-core shell model with continuum}",
    eprint = "1906.09258",
    archivePrefix = "arXiv",
    primaryClass = "nucl-th",
    reportNumber = "LLNL-JRNL-777984-DRAFT",
    doi = "10.1103/PhysRevC.100.024304",
    journal = "Phys. Rev. C",
    volume = "100",
    number = "2",
    pages = "024304",
    year = "2019"
}

@article{hebborn2022ab,
	title        = {Ab Initio Prediction of the He 4 (d, $\gamma$) Li 6 Big Bang Radiative Capture},
	author       = {Hebborn, Chlo{\"e} and Hupin, Guillaume and Kravvaris, Konstantinos and Quaglioni, Sofia and Navr{\'a}til, Petr and Gysbers, Peter},
	year         = 2022,
	journal      = {Physical Review Letters},
	publisher    = {APS},
	volume       = 129,
	number       = 4,
	pages        = {042503}
}

@article{TILLEY20023,
	title        = {Energy levels of light nuclei A=5, 6, 7},
	author       = {D.R. Tilley and C.M. Cheves and J.L. Godwin and G.M. Hale and H.M. Hofmann and J.H. Kelley and C.G. Sheu and H.R. Weller},
	year         = 2002,
	journal      = {Nuclear Physics A},
	volume       = 708,
	number       = 1,
	pages        = {3--163},
	doi          = {https://doi.org/10.1016/S0375-9474(02)00597-3},
	issn         = {0375-9474},
	url          = {https://www.sciencedirect.com/science/article/pii/S0375947402005973}
}

@misc{NIST:DLMF,
         key = "{\relax DLMF}",
       title = "{\it NIST Digital Library of Mathematical Functions}",
howpublished = "\url{https://dlmf.nist.gov/}, Release 1.2.3 of 2024-12-15",
         url = "https://dlmf.nist.gov/",
        note = "F.~W.~J. Olver, A.~B. {Olde Daalhuis}, D.~W. Lozier, B.~I. Schneider,
                R.~F. Boisvert, C.~W. Clark, B.~R. Miller, B.~V. Saunders,
                H.~S. Cohl, and M.~A. McClain, eds."}

@article{PhysRevC.5.391,
	title        = {Elastic Electron Scattering from $^{6}$Li and $^{7}$Li at Low Momentum Transfer},
	author       = {Bumiller, F. A. and Buskirk, F. R. and Dyer, J. N. and Monson, W. A.},
	year         = 1972,
	month        = {Feb},
	journal      = {Phys. Rev. C},
	publisher    = {American Physical Society},
	volume       = 5,
	pages        = {391--395},
	doi          = {10.1103/PhysRevC.5.391},
	url          = {https://link.aps.org/doi/10.1103/PhysRevC.5.391},
	issue        = 2,
	numpages     = {0}
}

@article{PhysRev.162.992,
	title        = {Elastic Electron Scattering from $^{6}$Li and $^{7}$Li},
	author       = {Suelzle, L. R. and Yearian, M. R. and Crannell, Hall},
	year         = 1967,
	month        = {Oct},
	journal      = {Phys. Rev.},
	publisher    = {American Physical Society},
	volume       = 162,
	pages        = {992--1005},
	doi          = {10.1103/PhysRev.162.992},
	url          = {https://link.aps.org/doi/10.1103/PhysRev.162.992},
	issue        = 4,
	numpages     = {0}
}

@article{VERMEER1989212,
title = {The value of B(E2; 32− → 12−) FOR 7Li},
journal = {Nuclear Physics A},
volume = {500},
number = {1},
pages = {212-220},
year = {1989},
issn = {0375-9474},
doi = {https://doi.org/10.1016/0375-9474(89)90138-3},
url = {https://www.sciencedirect.com/science/article/pii/0375947489901383},
author = {W.J. Vermeer and R.H. Spear and F.C. Barker}
}

@article{RUPAK2000405,
title = "Precision calculation of np$\rightarrow$d$\gamma$ cross section for big-bang nucleosynthesis",
journal = "Nuclear Physics A",
volume = "678",
number = "4",
pages = "405 - 423",
year = "2000",
issn = "0375-9474",
doi = "https://doi.org/10.1016/S0375-9474(00)00323-7",
url = "http://www.sciencedirect.com/science/article/pii/S0375947400003237",
author = "Gautam Rupak"
}

@article{ando2021cluster,
  title={Cluster effective field theory and nuclear reactions},
  author={Ando, Shung-Ichi},
  journal={The European Physical Journal A},
  volume={57},
  number={1},
  pages={17},
  year={2021},
  publisher={Springer}
}

@article{mukhamedzhanov2022asymptotic,
	title        = {Asymptotic normalization coefficients in nuclear reactions and nuclear astrophysics},
	author       = {Mukhamedzhanov, AM and Blokhintsev, LD},
	year         = 2022,
	journal      = {The European Physical Journal A},
	publisher    = {Springer},
	volume       = 58,
	number       = 2,
	pages        = 29
}

@article{BOUTEN1968385,
title = {Coulomb form factors of 7Li},
journal = {Nuclear Physics A},
volume = {111},
number = {2},
pages = {385-391},
year = {1968},
issn = {0375-9474},
doi = {https://doi.org/10.1016/0375-9474(68)90132-2},
url = {https://www.sciencedirect.com/science/article/pii/0375947468901322},
author = {M. Bouten and M.-C. Bouten and P. {Van Leuven}}
}

@article{Higa:2016igc,
	title        = {{Radiative$^{3}$He( $\alpha , \gamma$ )$^{7}$Be reaction in halo effective field theory}},
	author       = {Higa, Renato and Rupak, Gautam and Vaghani, Akshay},
	year         = 2018,
	journal      = {Eur. Phys. J. A},
	volume       = 54,
	number       = 5,
	pages        = 89,
	doi          = {10.1140/epja/i2018-12486-5},
	eprint       = {1612.08959},
	archiveprefix = {arXiv},
	primaryclass = {nucl-th}
}

@article{Poudel:2021mii,
	title        = {{Effective field theory analysis of $^{3}$He\textendash{}\ensuremath{\alpha} scattering data}},
	author       = {Poudel, Maheshwor and Phillips, Daniel R.},
	year         = 2022,
	journal      = {J. Phys. G},
	volume       = 49,
	number       = 4,
	pages        = {045102},
	doi          = {10.1088/1361-6471/ac4da6},
	note         = {[Erratum: J.Phys.G 49, 099601 (2022)]},
	eprint       = {2110.01451},
	archiveprefix = {arXiv},
	primaryclass = {nucl-th}
}

@article{Zhang:2019odg,
	title        = {{$S$-factor and scattering-parameter extractions from ${}^{3}\mathrm{He} +{}^{4}\mathrm{He} \rightarrow {}^{7}\mathrm{Be} + \gamma$}},
	author       = {Zhang, Xilin and Nollett, Kenneth M. and Phillips, D. R.},
	year         = 2020,
	journal      = {J. Phys. G},
	volume       = 47,
	number       = 5,
	pages        = {054002},
	doi          = {10.1088/1361-6471/ab6a71},
	eprint       = {1909.07287},
	archiveprefix = {arXiv},
	primaryclass = {nucl-th},
	reportnumber = {NT@UW-19-12 and INT-PUB-19-043}
}

@article{DESCOUVEMONT2004203,
	title        = {Compilation and R-matrix analysis of Big Bang nuclear reaction rates},
	author       = {Pierre Descouvemont and Abderrahim Adahchour and Carmen Angulo and Alain Coc and Elisabeth Vangioni-Flam},
	year         = 2004,
	journal      = {Atomic Data and Nuclear Data Tables},
	volume       = 88,
	number       = 1,
	pages        = {203--236},
	doi          = {https://doi.org/10.1016/j.adt.2004.08.001},
	issn         = {0092-640X},
	url          = {https://www.sciencedirect.com/science/article/pii/S0092640X04000282}
}

@article{PhysRevC.100.024304,
	title        = {$^{7}\mathrm{Be}$ and $^{7}\mathrm{Li}$ nuclei within the no-core shell model with continuum},
	author       = {Vorabbi, Matteo and Navr\'atil, Petr and Quaglioni, Sofia and Hupin, Guillaume},
	year         = 2019,
	month        = {Aug},
	journal      = {Phys. Rev. C},
	publisher    = {American Physical Society},
	volume       = 100,
	pages        = {024304},
	doi          = {10.1103/PhysRevC.100.024304},
	url          = {https://link.aps.org/doi/10.1103/PhysRevC.100.024304},
	issue        = 2,
	numpages     = 11
}

@article{Rand:1966zz,
	title        = {{Elastic Electron Scattering from the Magnetic Multipole Distributions of Li-6, Li-7, Be-9, B-10, B-11, and N-14}},
	author       = {Rand, R. E. and Frosch, R. and Yearian, M. R.},
	year         = 1966,
	journal      = {Phys. Rev.},
	volume       = 144,
	pages        = {859--873},
	doi          = {10.1103/PhysRev.144.859}
}

@article{Nollett:2001ub,
	title        = {{Radiative alpha capture cross-sections from realistic nucleon-nucleon interactions and variational Monte Carlo wave functions}},
	author       = {Nollett, Kenneth M.},
	year         = 2001,
	journal      = {Phys. Rev. C},
	volume       = 63,
	pages        = {054002},
	doi          = {10.1103/PhysRevC.63.054002},
	eprint       = {nucl-th/0102022},
	archiveprefix = {arXiv}
}

@article{Atkinson:2024zrm,
    author = "Atkinson, M. C. and Kravvaris, K. and Quaglioni, S. and Navr\'atil, P.",
    title = "{Ab initio calculation of the 3He(\ensuremath{\alpha},\ensuremath{\gamma})7Be astrophysical S factor with chiral two- and three-nucleon forces}",
    eprint = "2409.09515",
    archivePrefix = "arXiv",
    primaryClass = "nucl-th",
    doi = "10.1016/j.physletb.2024.139189",
    journal = "Phys. Lett. B",
    volume = "860",
    pages = "139189",
    year = "2025"
}

@article{Mertelmeier:1986egp,
    author = "Mertelmeier, T. and Hofmann, H. M.",
    title = "{Consistent cluster model description of the electromagnetic properties of lithium and beryllium nuclei}",
    doi = "10.1016/0375-9474(86)90141-7",
    journal = "Nucl. Phys. A",
    volume = "459",
    pages = "387--416",
    year = "1986"
}

@article{Dohet-Eraly:2015ooa,
    author = "Dohet-Eraly, J\'er\'emy and Navr\'atil, Petr and Quaglioni, Sofia and Horiuchi, Wataru and Hupin, Guillaume and Raimondi, Francesco",
    title = "{$^3$He($\alpha,\gamma$)$^7$Be and $^3$H($\alpha,\gamma$)$^7$Li astrophysical S factors from the no-core shell model with continuum}",
    eprint = "1510.07717",
    archivePrefix = "arXiv",
    primaryClass = "nucl-th",
    reportNumber = "LLNL-JRNL-678557",
    doi = "10.1016/j.physletb.2016.04.021",
    journal = "Phys. Lett. B",
    volume = "757",
    pages = "430--436",
    year = "2016"
}

@article{PhysRevC.85.045807,
      title = "{Determination of the He-3 + alpha ---\ensuremath{>} Be-7 asymptotic normalization coefficients (nuclear vertex constants) and their application for extrapolation of the He-3(alpha,gamma)Be-7 astrophysical S-factors to the solar energy region}",
  author = {Tursunmahatov, Q. I. and Yarmukhamedov, R.},
  journal = {Phys. Rev. C},
  volume = {85},
  issue = {4},
  pages = {045807},
  numpages = {14},
  year = {2012},
  month = {Apr},
  publisher = {American Physical Society},
  doi = {10.1103/PhysRevC.85.045807},
  url = {https://link.aps.org/doi/10.1103/PhysRevC.85.045807}
}

@article{kajino1984electromagnetic,
	title        = {Electromagnetic properties of 7Li and 7Be in a cluster model},
	author       = {Kajino, Toshitaka and Matsuse, Takehiro and Arima, Akito},
	year         = 1984,
	journal      = {Nuclear Physics A},
	publisher    = {Elsevier},
	volume       = 413,
	number       = 2,
	pages        = {323--352}
}

@article{kajino19863he,
  title={The 3He ($\alpha$, $\gamma$) 7Be and 3He ($\alpha$, $\gamma$) 7Li reactions at astrophysical energies},
  author={Kajino, Toshitaka},
  journal={Nuclear Physics A},
  volume={460},
  number={3},
  pages={559--580},
  year={1986},
  publisher={Elsevier}
}

@article{RevModPhys.88.015004,
  title = {Big bang nucleosynthesis: Present status},
  author = {Cyburt, Richard H. and Fields, Brian D. and Olive, Keith A. and Yeh, Tsung-Han},
  journal = {Rev. Mod. Phys.},
  volume = {88},
  issue = {1},
  pages = {015004},
  numpages = {22},
  year = {2016},
  month = {Feb},
  publisher = {American Physical Society},
  doi = {10.1103/RevModPhys.88.015004},
  url = {https://link.aps.org/doi/10.1103/RevModPhys.88.015004}
}

@article{Asplund:2005yt,
    author = "Asplund, Martin and Lambert, David L. and Nissen, Poul Erik and Primas, Francesca and Smith, Verne V.",
    title = "{Lithium isotopic abundances in metal-poor halo stars}",
    eprint = "astro-ph/0510636",
    archivePrefix = "arXiv",
    doi = "10.1086/503538",
    journal = "Astrophys. J.",
    volume = "644",
    pages = "229--259",
    year = "2006"
}

@article{osti_1802279,
title = {Big-Bang Nucleosynthesis after Planck},
author = {Fields, Brian D. and Olive, Keith A. and Yeh, Tsung-Han and Young, Charles},
abstractNote = {We assess the status of big-bang nucleosynthesis (BBN) in light of the final Planck data release and other recent developments, and in anticipation of future measurements. Planck data from the recombination era fix the cosmic baryon density to 0.9% precision, and now damping tail measurements determine the helium abundance and effective number of neutrinos with precision approaching that of astronomical and BBN determinations respectively. All three parameters are related by BBN. In addition, new high-redshift measurements give D/H to better precision than theoretical predictions, and new Li/H data reconfirm the lithium problem. We present new 7Be(n, p) 7Li rates using new neutron capture measurements; we have also examined the effect of proposed changes in the d(p, γ) 3He rates. Using these results we perform a series of likelihood analyses. We assess BBN/CMB consistency, with attention to how our results depend on the choice of Planck data, as well as how the results depend on the choice of non-BBN, non-Planck data sets. Most importantly the lithium problem remains, and indeed is more acute given the very tight D/H observational constraints; new neutron capture data reveals systematics that somewhat increases uncertainty and thus slightly reduces but does not essentially change the problem. We confirm that d(p, γ) 3He theoretical rates brings D/H out of agreement and slightly increases 7Li; new experimental data are needed at BBN energies. Setting the lithium problem aside, we find the effective number of neutrino species at BBN is Nν = 2.86±0.15. Future CMB Stage-4 measurements promise substantial improvements in BBN parameters: helium abundance determinations will be competitive with the best astronomical determinations, and Neff will approach sensitivities capable of detecting the effects of Standard Model neutrino heating of the primordial plasma.},
doi = {10.1088/1475-7516/2020/03/010},
journal = {Journal of Cosmology and Astroparticle Physics},
number = 03,
volume = 2020,
place = {United States},
year = {2020},
month = {3}
}

@article{Nortershauser:2008vp,
    author = "Nortershauser, W. and others",
    title = "{Nuclear Charge Radii of Be-7, Be-9, Be-10 and the one-neutron halo nucleus Be-11}",
    eprint = "0809.2607",
    archivePrefix = "arXiv",
    primaryClass = "nucl-ex",
    doi = "10.1103/PhysRevLett.102.062503",
    journal = "Phys. Rev. Lett.",
    volume = "102",
    pages = "062503",
    year = "2009"
}

@article{van1971magnetization,
	title        = {Magnetization distribution of the 7Li nucleus as obtained from electron scattering through 180°. The electric quadrupole moment of 7Li},
	author       = {Van Niftrik, GJC and Lapik{\'a}s, La and De Vries, H and Box, G},
	year         = 1971,
	journal      = {Nuclear Physics A},
	publisher    = {Elsevier},
	volume       = 174,
	number       = 1,
	pages        = {173--192}
}

@article{ANGELI201369,
title = {Table of experimental nuclear ground state charge radii: An update},
journal = {Atomic Data and Nuclear Data Tables},
volume = {99},
number = {1},
pages = {69-95},
year = {2013},
issn = {0092-640X},
doi = {https://doi.org/10.1016/j.adt.2011.12.006},
url = {https://www.sciencedirect.com/science/article/pii/S0092640X12000265},
author = {I. Angeli and K.P. Marinova}
}

@article{buck1985cluster,
	title        = {Cluster model of A= 7 nuclei and the astrophysical S factor for 3He ($\alpha$, $\gamma$) 7Be at zero energy},
	author       = {Buck, B and Baldock, RA and Rubio, JA},
	year         = 1985,
	journal      = {Journal of Physics G: Nuclear Physics},
	publisher    = {IOP Publishing},
	volume       = 11,
	number       = 1,
	pages        = {L11}
}

@article{AMROUN1994596,
	title        = {3H and 3He electromagnetic form factors},
	author       = {A. Amroun and V. Breton and J.-M. Cavedon and B. Frois and D. Goutte and F.P. Juster and Ph. Leconte and J. Martino and Y. Mizuno and X.-H. Phan and S.K. Platchkov and I. Sick and S. Williamson},
	year         = 1994,
	journal      = {Nuclear Physics A},
	volume       = 579,
	number       = 3,
	pages        = {596--626},
	doi          = {https://doi.org/10.1016/0375-9474(94)90925-3},
	issn         = {0375-9474},
	url          = {https://www.sciencedirect.com/science/article/pii/0375947494909253}
}

@article{STONE20161,
	title        = {Table of nuclear electric quadrupole moments},
	author       = {N.J. Stone},
	year         = 2016,
	journal      = {Atomic Data and Nuclear Data Tables},
	volume       = {111-112},
	pages        = {1--28},
	doi          = {https://doi.org/10.1016/j.adt.2015.12.002},
	issn         = {0092-640X},
	url          = {https://www.sciencedirect.com/science/article/pii/S0092640X16000024}
}

@book{Ericson:1988gk,
	title        = {{Pions and Nuclei}},
	author       = {Ericson, Torleif Erik Oskar and Weise, W.},
	year         = 1988,
	publisher    = {Clarendon Press},
	address      = {Oxford, UK},
	isbn         = {978-0-19-852008-5}
}

@article{Jones:1972ky,
	title        = {{Multipole gamma N Delta form-factors and resonant photoproduction and electroproduction}},
	author       = {Jones, H. F. and Scadron, M. D.},
	year         = 1973,
	journal      = {Annals Phys.},
	volume       = 81,
	pages        = {1--14},
	doi          = {10.1016/0003-4916(73)90476-4}
}

@article{PhysRevC.60.065205,
  title = {$n\stackrel{\ensuremath{\rightarrow}}{p}d\ensuremath{\gamma}$ for big-bang nucleosynthesis},
  author = {Chen, Jiunn-Wei and Savage, Martin J.},
  journal = {Phys. Rev. C},
  volume = {60},
  issue = {6},
  pages = {065205},
  numpages = {6},
  year = {1999},
  month = {Nov},
  publisher = {American Physical Society},
  doi = {10.1103/PhysRevC.60.065205},
  url = {https://link.aps.org/doi/10.1103/PhysRevC.60.065205}
}

@article{Ryberg:2014exa,
    author = "Ryberg, Emil and Forss{\'e}n, Christian and Hammer, H. -W. and Platter, Lucas",
    title = "{Constraining Low-Energy Proton Capture on Beryllium-7 through Charge Radius Measurements}",
    eprint = "1406.6908",
    archivePrefix = "arXiv",
    primaryClass = "nucl-th",
    doi = "10.1140/epja/i2014-14170-2",
    journal = "Eur. Phys. J. A",
    volume = "50",
    pages = "170",
    year = "2014"
}

@article{Siegert:1937yt,
    author = "Siegert, A. J. F.",
    title = "{Note on the interaction between nuclei and electromagnetic radiation}",
    doi = "10.1103/PhysRev.52.787",
    journal = "Phys. Rev.",
    volume = "52",
    pages = "787--789",
    year = "1937"
}

@article{Pervin:2007sc,
    author = "Pervin, Muslema and Pieper, Steven C. and Wiringa, Robert B.",
    title = "{Quantum Monte Carlo calculations of electroweak transition matrix elements in A = 6,7 nuclei}",
    eprint = "0710.1265",
    archivePrefix = "arXiv",
    primaryClass = "nucl-th",
    doi = "10.1103/PhysRevC.76.064319",
    journal = "Phys. Rev. C",
    volume = "76",
    pages = "064319",
    year = "2007"
}

@article{PhysRevC.87.035503,
  title = {Quantum Monte Carlo calculations of electromagnetic moments and transitions in $A\ensuremath{\le}9$ nuclei with meson-exchange currents derived from chiral effective field theory},
  author = {Pastore, S. and Pieper, Steven C. and Schiavilla, R. and Wiringa, R. B.},
  journal = {Phys. Rev. C},
  volume = {87},
  issue = {3},
  pages = {035503},
  numpages = {15},
  year = {2013},
  month = {Mar},
  publisher = {American Physical Society},
  doi = {10.1103/PhysRevC.87.035503},
  url = {https://link.aps.org/doi/10.1103/PhysRevC.87.035503}
}

@misc{claude-opus-5,
  author       = {{Anthropic}},
  title        = {Claude Opus 5},
  year         = {2026},
  howpublished = {\url{https://claude.ai}},
  note         = {Large language model. Accessed August 2026}
}

@misc{ChatGPT,
  author       = {{OpenAI}},
  title        = {{ChatGPT}},
  howpublished = {\url{https://chat.openai.com}},
  note         = {Large language model, accessed 2024--2025},
  year         = {2025}
}

\end{document}